\documentclass{aastex6}
\usepackage{mkfig}

\begin{document}

\title{Evidence for Asymmetry in the Velocity Distribution of the
  Interstellar Neutral Helium Flow Observed by {\em IBEX} and {\em Ulysses}}

\author{Brian E. Wood\altaffilmark{1}, Hans-Reinhard
  M\"{u}ller\altaffilmark{2}, Eberhard M\"{o}bius\altaffilmark{3}}
\altaffiltext{1}{Naval Research Laboratory, Space Science Division,
  Washington, DC 20375, USA; brian.wood@nrl.navy.mil}
\altaffiltext{2}{Department of Physics and Astronomy, Dartmouth College,
  Hanover, NH 03755, USA}
\altaffiltext{3}{Space Science Center and Department of Physics,
  University of New Hampshire, Durham, NH 03824 USA}


\begin{abstract}

     We use observations from the {\em Interstellar Boundary Explorer}
(IBEX) and {\em Ulysses} to explore the possibility that the
interstellar neutral helium flowing through the inner solar system
possesses an intrinsic non-Maxwellian velocity distribution.  In
fitting the {\em IBEX} and {\em Ulysses} data, we experiment with
both a kappa distribution and a bi-Maxwellian, instead of the
usual Maxwellian assumption.  The kappa distribution does not improve
the quality of fit to either the {\em IBEX} or {\em Ulysses} data,
and we find lower limits to the kappa parameter of $\kappa>12.1$ and
$\kappa>6.0$ from the {\em IBEX} and {\em Ulysses} analyses, respectively.
In contrast, we do find evidence that a bi-Maxwellian improves fit
quality.  For {\em IBEX}, there is a clear preferred
bi-Maxwellian solution with $T_{\perp}/T_{\parallel}=0.62\pm 0.11$
oriented about an axis direction with ecliptic coordinates
$(\lambda_{axis},b_{axis})=(57.2^{\circ}\pm 8.9^{\circ},
-1.6^{\circ}\pm 5.9^{\circ})$.
The {\em Ulysses} data provide support for this result, albeit with
lower statistical significance.  The axis direction is close
to the ISM flow direction, in a heliocentric rest frame, and is 
therefore unlikely to be indicative of velocity distribution
asymmetries intrinsic to the ISM.  It is far more likely that
these results indicate the presence of asymmetries induced
by interactions in the outer heliosphere.

\end{abstract}



\section{Introduction}

     The global characteristics of the heliosphere surrounding the
Sun are defined in large part by the characteristics of the
surrounding very local interstellar medium (VLISM).  The plasma component
of the VLISM is deflected around the heliosphere at the heliopause and
has therefore been unavailable for direct observation, at least
until recently with the {\em Voyager} 1 and 2 spacecraft crossing
the heliopause in 2012 and 2018, respectively \citep{ecs13,dag13,ecs18}.
Much of what we know about the
VLISM comes instead from interstellar
neutrals (ISNs), which can penetrate the heliopause and therefore be
observed more locally in the inner solar system at or near 1~AU.

     The first studies of the local ISN flow
involved the analysis of solar Lyman-$\alpha$ backscatter
from neutral H \citep{jlb71,eq99}.
However, better measurements of VLISM properties
come not from neutral H but from neutral He.  This is because
charge exchange processes significantly alter the characteristics of
neutral H as it travels through the heliosphere.  In contrast,
He has much lower charge exchange rates and therefore
reaches the inner solar system in a more pristine state.

     Somewhat analogous to H, the first detections of interstellar
neutral He in the inner solar system were radiative in nature,
specifically scattered solar He~I $\lambda$584 emission from the He
focusing cone downwind from the Sun \citep{rrm72}.  However,
the most precise measurements of the neutral He properties have
come from particle detectors that can study the He flow directly.
The GAS instrument on the {\em Ulysses} spacecraft was the first
instrument capable of these measurements \citep{mw04,bew15}.
The ongoing {\em Interstellar Boundary
Explorer} (IBEX) mission continues this legacy, specifically with
the IBEX-Lo instrument \citep{em12,mb15}.

     Analysis of these He measurements requires a forward
modeling exercise, where ISNs, with a velocity distribution
function (VDF) defined by some set of parameters, are propagated into
the inner solar system under the influence of primarily the Sun's
gravity, and secondarily various loss processes, particularly
photoionization close to the Sun.  The parameters of the assumed VDF
are varied to maximize agreement with the data.  The usual assumption
in such studies is to assume that the VDF is a Maxwellian far
from the Sun.

     One indication that the Maxwellian assumption may be
overly simplistic is the discovery of a secondary neutral He component
\citep{mak14,mak16}.  This component is very clear in {\em IBEX}
data, but hints of it have also been found in the noisier {\em
Ulysses} data \citep{bew17}.  The secondary component is likely
created by charge exchange processes in the outer heliosheath beyond
the heliopause.  Its existence therefore demonstrates that despite low
charge exchange rates, even He undergoes some level of
interaction in the outer heliosphere, which might affect the primary
He flow properties as well.

     Even in the absence of heliospheric interaction, it is far from
certain that the pristine ISN He VDF is a pure Maxwellian.
Measurements of interstellar absorption lines towards nearby stars
indicate the presence of substantial nonthermal velocities in the
nearby ISM, generally interpreted as turbulence of some sort \citep{sr08},
but which could also be indicative of velocity
gradients in the ISM \citep{cg14}.  In the presence of
turbulence, it is possible that a kappa distribution may more
accurately describe the ISN VDF than a Maxwellian \citep{laf06,gl18}.
Furthermore, there is a significant magnetic
field in the VLISM, and MHD wave/particle interactions could lead to
the plasma component being bi-Maxwellian in nature, with different
effective temperatures parallel and perpendicular to the magnetic
field.  Charge exchange processes could imprint such plasma properties
onto the neutrals.  \citet{srs11} collectively used
measurements of nearby interstellar temperatures and nonthermal
velocities to search for evidence of anisotropies indicative of such
MHD processes.  They found none, but the quoted limit of
$T_{\perp}/T_{\parallel}<1.67$ still allows for the possibility of
significant undetected anisotropy.

     We here present a reanalysis of {\em IBEX} and {\em Ulysses} data,
experimenting with both kappa and bi-Maxwellian distributions.  The
goal is to assess whether such assumptions demonstrably lead to
better fits to the data than the usual Maxwellian assumption.  We
will also assess whether assuming a non-Maxwellian VDF far from the
Sun in any way affects the flow velocity and direction inferred
from the data.

\section{Assuming a Bi-Maxwellian Distribution}

     The usual assumption for the VLISM He VDF far from
the Sun is a simple Maxwellian,
\begin{equation}
F({\bf v})=n \left(\frac{m}{2\pi kT}\right)^{\frac{3}{2}}
  \exp\left(-\frac{m ({\bf v}-{\bf U})^2}{2kT}\right),
\end{equation}
where $m$ is the mass of the He atom, and $k$ the Boltzmann
constant.  The flow parameters of
interest are the temperature, $T$, the He density
$n$, and the flow vector ${\bf U}=(V_{flow},\lambda_{flow},\beta_{flow})$;
where $V_{flow}$ is the flow magnitude, and $\lambda_{flow}$
and $\beta_{flow}$ are the flow longitude and latitude,
respectively, in ecliptic coordinates.

     However, we here replace the Maxwellian with a bi-Maxwellian VDF,
\begin{equation}
F({\bf v})=\frac{n}{T_{\perp}T_{\parallel}^{1/2}}
  \left(\frac{m}{2\pi k}\right)^{\frac{3}{2}}
  \exp\left[-\frac{m}{2k}\left(
  \frac{({\bf v_{\parallel}}-{\bf U_{\parallel}})^2}{T_{\parallel}}+
  \frac{({\bf v_{\perp}}-{\bf U_{\perp}})^2}{T_{\perp}}\right)\right],
\end{equation}
where the velocities ${\bf v}$ and ${\bf U}$ have been decomposed into
components perpendicular (${\bf v_{\perp}}$, ${\bf U_{\perp}}$) and
parallel (${\bf v_{\parallel}}$, ${\bf U_{\parallel}}$)
to some axis direction, the axis being defined by a direction in
ecliptic coordinates, $(\lambda_{axis},\beta_{axis})$.  Different
temperatures are assumed perpendicular ($T_{\perp}$) and parallel
($T_{\parallel}$) to the axis.  The five parameters of the
Maxwellian increase to eight parameters for
the bi-Maxwellian ($n$, $V_{flow}$, $\lambda_{flow}$, $\beta_{flow}$,
$T_{\perp}$, $T_{\parallel}$, $\lambda_{axis}$, $\beta_{axis}$).

     A bi-Maxwellian assumption makes the most physical sense if
the axis direction is defined by the magnetic field, since MHD
processes might be expected to produce a bi-Maxwellian VDF ordered
about the field direction.  However, a bi-Maxwellian can also be
considered to be a generic parametrized approximation of any
asymmetric VDF.  In that spirit, we will not be forcing the axis to be
in the VLISM field direction.

\subsection{{\em IBEX} Data Analysis}

     We first describe our bi-Maxwellian analysis of {\em IBEX} data.
The {\em IBEX} spacecraft was launched into a highly elliptical orbit around
Earth on 2008 October 19, and has been accumulating data since that
time.  The goal of the mission is to generate maps of various
populations of neutrals coming from the outer heliosphere and the VLISM
\citep{djm09}.  The spacecraft operates in a continuous spin,
with the spin axis pointed very near the Sun, and with {\em IBEX}'s two
instruments, IBEX-Lo and IBEX-Hi, observing radially away from the
spin axis.  Observations are therefore made along a
$7^{\circ}$ wide latitudinal strip that is roughly normal to the
Sun-spacecraft line.  The spin axis is maintained in a steady
orientation in an inertial rest frame for a time period of many days,
during which the spin axis drifts wesward of the Sun due to the Earth's
orbital motion.  The spin axis is periodically adjusted to reorient
it towards the Sun.  In this way, {\em IBEX} gradually scans across the
entire sky once every six months.

     The low energy interstellar neutrals are studied using
the IBEX-Lo instrument \citep{sf09}.  Although its spin-oriented
design allows the sky to be mapped every six months, in practice the
interstellar neutrals can only be observed when the orbital motion
around the Sun is moving the spacecraft towards the upwind direction
of the flow, which only happens once per year.  The IBEX-Lo instrument
is capable of detecting a number of different ISN species, but we will
be focusing only on He here.  The ISN He observing season is in the
months of January and February each year, during which {\em IBEX}
gradually scans across the He beam \citep{em09}.

\begin{figure}[t]
\plotfiddle{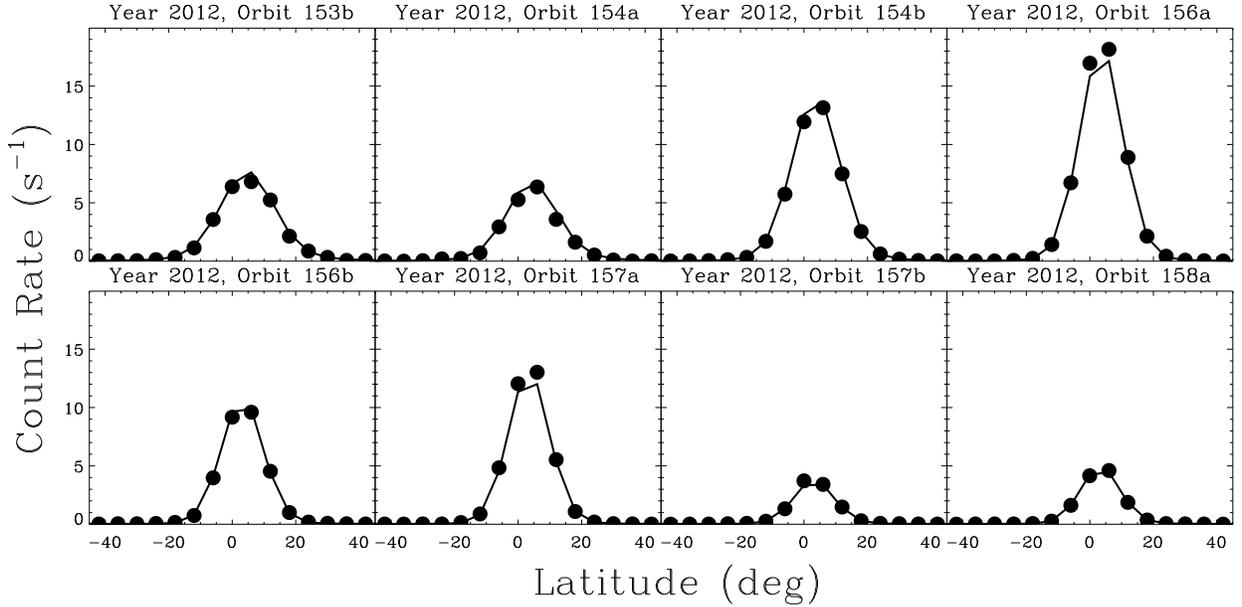}{3.2in}{90}{90}{90}{370}{-260}
\caption{The ISN He count rate observed by {\em IBEX} as a function of
  ecliptic latitude for the sequence of orbits in which {\em IBEX} scans
  across the ISN He beam in 2012.  The solid line is a fit to the
  data assuming a Maxwellian VDF.}
\end{figure}
     The {\em IBEX} observations are typically organized by sequential
Earth orbit number.  The {\em IBEX} spacecraft initially orbited Earth once
every 7.6~days, but in mid-2011 {\em IBEX} was shifted to a more stable
orbit with an orbital period closer to 9 days.  The data are carefully
screened to consider only data free from various sources of
contamination \citep{em12,em15}.  Figure~1 in \citet{mb15}
provides a useful illustration of the good time intervals
available in each orbit.  Figure~1 (in this paper) shows a sequence of
He count rates as a function of latitude observed during the orbits of
the 2012 ISN He observing season.  [See \citet{mb15} for
depictions of count rates from other years.]
Originally, the {\em IBEX} spin axis was
adjusted only once per orbit, but since the 2011 orbit adjustment the
spin axis has been reoriented towards the Sun twice per orbit; hence
the ``a'' and ``b'' appendages to the orbit numbers in Figure~1.  The
count rates actually vary significantly within each orbit, due to the
ever changing pointing direction relative to the He beam, induced by
the Earth's orbital motion; and also due to {\em IBEX}'s own
orbital motion around the Earth.  The count rates displayed are
averages over the full orbital time periods.

     Figure~1 illustrates how the He count rates increase
with orbit as the {\em IBEX} scan direction moves closer to the center
of the He beam, and then decrease thereafter.  The pattern is
not as smooth as might be expected, due in part to irregularities
in the periodic spin axis adjustments, and also due to the data
screening described above, which leads to the actual observing
times varying significantly from orbit to orbit.  A substantial
fraction of the total observing time is excluded by the screening.
Orbit 155 is missing entirely from Figure~1, with there being no
acceptable observing periods during that orbit.

     Our data selection philosophy mirrors that of \citet{mb15},
with the general idea being to focus only on the
orbits with the highest fluxes from the primary He beam, where the
effects of minor contaminating sources, such as the secondary neutral
population mentioned in Section~1, should be minor.  Figure~1 shows the
only orbits considered from 2012.  Compared to \citet{mb15},
we consider a slightly broader latitudinal range of $\pm
25^{\circ}$ from the ecliptic.

     Fitting the data requires a forward modeling exercise of
propagating particles with an assumed initial VDF towards the Sun
under the influence of the Sun's gravity.  Our particle tracking code,
which has been used in past analyses of {\em Ulysses} data
\citep{bew15,bew17}, and which is described in detail
elsewhere \citep{hrm12a,hrm12b,hrm13},
takes advantage of certain conservation
properties for particles moving in Keplerian orbits to map a VDF
at infinity to one close to the Sun.  This approach is
significantly different from that of \citet{mb15}, which
is described in detail by \citet{jms15}.

     For {\em IBEX}, it is important to consider the effects of loss
processes, particularly photoionization.  The Earth's orbital position
changes significantly during the two months it takes to scan across
the He beam, which means exposure to photoionization changes
significantly.  In particular, earlier orbits (e.g., orbit 153b in
Figure~1) suffer larger losses than later orbits (e.g., orbit 158a).
Thus, the loss processes will affect how the overall flux level
changes with orbit, which must be corrected for in the analysis.  In
principle, the loss rates will be both time and latitude dependent, in
a way that is difficult to precisely model.  We do not attempt to
account for this dependence in any great detail.  Instead, we use the
loss rates estimated by \citet{jms15} to compute the average
rate during the observing season in question, and that single value is
assumed in the particle tracking for that entire season, and for
particles coming from all directions.  For 2009--2014, the
assumed loss rates at 1~AU in units of $10^{-8}$ s$^{-1}$ are
0.73, 0.84, 0.95, 1.18, 1.22, and 1.44, respectively.
The actual time variability
of the loss rate within an observing season is estimated to be
no more than $\sim 10$\%.

\begin{figure}[t]
\plotfiddle{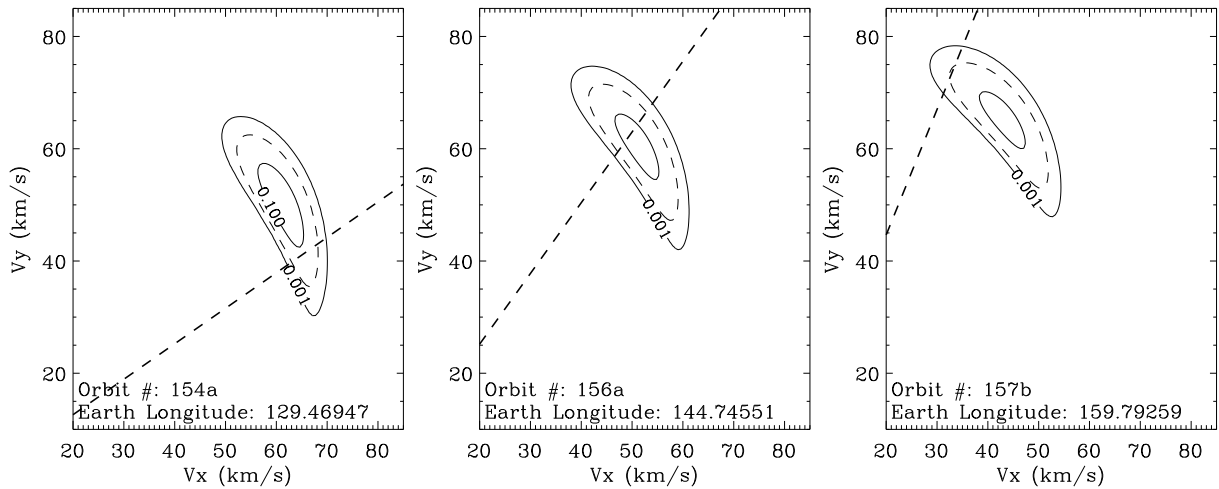}{2.5in}{90}{70}{70}{280}{-190}
\caption{An illustration of how {\em IBEX} scans across the ISN He beam
  each year.  Contours of the ISN He VDF in the $V_z=0$ km~s$^{-1}$
  plane are shown, at the location of Earth/{\em IBEX} and in the spacecraft
  rest frame, for three different Earth longitudinal locations (in a
  heliocentric Aries ecliptic coordinate system).  The dashed lines are
  the scan directions of {\em IBEX} through the VDF in the ecliptic plane,
  using the spin axis directions of the 2012 observing season, defined
  by the orbit numbers provided in the figure.
  A movie version of this figure is available online.  In the movie
  version the effects of the spacecraft orbital motion around Earth
  are easily distinguishable from the effects of
  Earth's motion around the Sun.}
\end{figure}
     The particle tracking code provides a model VDF at {\em IBEX}'s
location at any time and location, which when put into the spacecraft
rest frame can then be used to infer model count rates.  Figure~2
shows examples of VDFs at {\em IBEX}'s location in the spacecraft frame, or
at least slices through the three-dimensional VDFs for the $V_z=0$
km~s$^{-1}$ plane appropriate for viewing directions near the
ecliptic.  The coordinate system used is a heliocentric Aries ecliptic
coordinate system, with the z-axis towards ecliptic north, and the x-axis
towards the vernal equinox (i.e., the first point of Aries).
Dashed lines show {\em IBEX}'s viewing angle through the VDF in
the ecliptic, illustrating how {\em IBEX} gradually scans across the VDF
over the course of many orbits.

     In order to compare with the data as displayed in Figure~1,
it is necessary to integrate over the full orbital period to model
the displayed average count rates over this period.  The particle
tracking computations are numerically intensive, which places
practical limits on the time resolution for which VDFs can be
computed.  We ultimately determine that a time resolution of one day
is sufficient for our purposes.  Using a set of VDFs with this time
resolution, we can compute the observed count rate as a function of
time with this temporal resolution, and then compute the average rate
within the orbit for comparison with the data as displayed in
Figure~1, naturally taking into account the exact time intervals
utilized within the orbit.

     The calculation considers the collimator transmission
function of IBEX-Lo \citep{sf09,mb12,jms15}, which describes
how the instrument response
varies as a function of angular location relative to the central
pointing direction.  The exact response curve is not purely a function
of angular distance from the centerpoint, but has an azimuthal
component with a hexagonal character.  For simplicity, we average over
this azimuthal dependence in order to reduce the collimator response
to a simple dependence on angle from the centerpoint.

     Finally, the calculation also includes a correction for
latitudinal scan smoothing.  In Figure~1, the counts collected by
{\em IBEX} are placed into $6^{\circ}$ latitude bins, but due to {\em IBEX}'s
spinning nature the exposure time is distributed uniformly throughout
the $\pm 3^{\circ}$ range of each bin, which will smooth the curves
in Figure~1 somewhat.  We account for this in our model by computing
count rates for ten sub-bins within each $6^{\circ}$ latitude bin,
and then use the average of the sub-bin rates as the model rate for
that bin.

     Initially, we consider the first six years of {\em IBEX} data in
our analysis, from 2009-2014, limiting ourselves to the orbits
where He count rates are high, as in \citet{mb15}.
Although the secondary He component is comparatively weak for these
orbits, we correct for it assuming the secondary He flow parameters of
\citet{mak16}.  From those parameters, the model secondary He
count rates are derived using our particle tracking code, and then
subtracted from the data before doing a data/model comparison for the
primary He component.  In addition to the flow parameters of interest,
there is a normalization factor and a flat background count rate that
are additional free parameters of the fit.  These are allowed to be
different for each year's data, so this adds up to an additional
twelve free parameters if six years of data are considered.

     The normalization factors are redundant with the density
parameter, $n$, with $n$ also acting as a simple
multiplicative factor [see Equations (1) and (2)].  If the He
detection efficiency was known well, the normalization factors would
provide an $n$ measurement for each year.  However, the detection
efficiencies are not well quantified, due to uncertainties in the
absolute calibration of the instrument with respect to He detection.
Interstellar densities are very unlikely to change from year to
year, but we allow the normalization factors to vary each year to
account for possible changes in detector sensitivity with time.
For noble gases like He, the IBEX-Lo detector detects He via sputtering,
where the ISN neutrals produce various ionized sputtering products
within the IBEX-Lo detector and it is these ions that are actually
being observed.

     We use the $\chi^2$ parameter as our quality-of-fit indicator
\citep{prb92}, so the fitting process involves
finding the set of parameters that minimizes $\chi^2$.  If $\nu$
is the number of degrees of freedom of the fit (the number of
data points minus the number of free parameters), then the reduced
chi-squared is defined as $\chi^2_{\nu}=\chi^2/\nu$, which should be
$\sim 1$ for a good fit.  The $\chi^2$ minimization is performed using
the Marquardt method \citep{whp89}.  This involves an
initial guess for the fit parameters, a determination of the
$\chi^2$ value associated with that guess, and then an iterative
process of changing the parameters to follow the $\chi^2$ gradient
down to its minimum value in parameter space.

     Our first fits to the {\em IBEX} data simply assume a Maxwellian VDF.
Considering all 2009--2014 data,
we initially find best-fit flow parameters that are not in very
good agreement with canonical {\em Ulysses} measurements, with a low
velocity of $V_{flow}=24.2$ km~s$^{-1}$ and a high flow
longitude of $\lambda_{flow}=77.4^{\circ}$.
However, these discrepancies disappear when the 2009--2010 data
are ignored, consistent with the results of \citet{mb15}, who also find
lower $V_{flow}$ and higher $\lambda_{flow}$ for 2009--2010.
Furthermore, ignoring the 2009--2010 data also improves our quality of
fit overall.

\begin{deluxetable}{lccc}
\tabletypesize{\scriptsize}
\tablecaption{Recent Interstellar He Flow Measurements}
\tablecolumns{4}
\tablewidth{0pt}
\tablehead{
  \colhead{} &\colhead{Wood et al.\ (2015)}&\colhead{Bzowski et al.\ (2015)} &
    \colhead{This paper}}
\startdata
Data Source              & {\em Ulysses} & {\em IBEX}    & {\em IBEX} \\
Assumed VDF              & Maxwellian    & Maxwellian    & bi-Maxwellian \\
$V_{flow}$ (km~s$^{-1}$) &$26.08\pm 0.21$& $25.8\pm 0.4$ & $25.6\pm 1.2$ \\
$\lambda_{flow}$ (deg)   &$75.54\pm 0.19$& $75.8\pm 0.5$ & $75.6\pm 1.5$ \\
$b_{flow}$ (deg)         &$-5.44\pm 0.24$&$-5.16\pm 0.10$&$-5.09\pm 0.14$\\
$T$ (K)                  & $7260\pm 270$ & $7440\pm 260$ &     ...       \\
$T_{\perp}$ (K)          &      ...      &      ...      & $7580\pm 960$ \\
$T_{\parallel}$ (K)      &      ...      &      ...      &$12700\pm 2960$\\
$T_{\perp}/T_{\parallel}$&      ...      &      ...      & $0.62\pm 0.11$\\
$\lambda_{axis}$ (deg)   &      ...      &      ...      & $57.2\pm 8.9$ \\
$b_{axis}$ (deg)         &      ...      &      ...      & $-1.6\pm 5.9$ \\
\enddata
\end{deluxetable}
   Thus, we ultimately decide to exclude the 2009--2010
data from our analysis, considering only the four years of 2011--2014.
Based on the 2011--2014 data, our best Maxwellian fit
parameters become $V_{flow}=25.4$ km~s$^{-1}$,
$(\lambda_{flow},b_{flow})=(75.7^{\circ},-5.1^{\circ})$, and $T=7700$~K.
Table~1 shows that these results are in reasonable agreement with our
past {\em Ulysses} measurements \citep{bew15}, and the {\em IBEX}
measurements of \citet{mb15}.  With regards to the quoted uncertainties,
it is worth noting that in this paper and in \citet{bew15} we use
3$\sigma$ uncertainties, while \citet{mb15} quote 1$\sigma$ errors.
It is not entirely clear why the 2009--2010 data seem problematic, but
one possibility is contamination from interstellar H and secondary He.
The IBEX-Lo detector is capable of observing ISN H, which could in
principle be a contaminant in the He count rates \citep{em09,ps18}.
Furthermore, the H contamination would have been stronger in the solar
minimum period of 2009--2010 than in subsequent more active years,
with photoionization losses greatly reducing H fluxes
\citep{ls13,ag19}.

     The fit is explicitly shown in Figure~1, at least for the 2012 data.
There are significant discrepancies with the data, discrepancies that
are essentially identical to those seen in the \citet{mb15}
fits.  For example, the
underprediction of counts for orbits 156a and 157a is also apparent in
Figure~6 of \citet{mb15}.  This consistency implies that the
models are not an issue here, but unknown systematic uncertainties in the
data are instead the cause.  As a consequence of these uncertainties, the
best-fit $\chi^2_{\nu}$ value of the fit is higher than it should be,
specifically $\chi^2_{\nu}=2.28$, with the number of degrees of
freedom for the fit being 291.  \citet{ps15} provide an
extensive discussion of many possible sources of systematic error in
{\em IBEX} data.

     One of the fundamental issues that plagues analysis of {\em IBEX}
data is parameter degeneracy, which means that the fit parameters
are highly dependent on each other.  Among the strongest dependencies
is the one between $V_{flow}$ and $\lambda_{flow}$.  This can be
understood by considering the following question:  If you see the
He beam coming from a given direction, does that imply a fast He
flow coming from that direction, or is it a slower He flow being
deflected into that direction by the Sun's gravity?  Since the VLISM
flow is roughly in the ecliptic plane where {\em IBEX} is operating, the
issue mostly expresses itself as a correlation between velocity
and longitude rather than latitude.  Using a sequence of
single-Maxwellian fits, we determine the best-fit velocity as a
function of assumed flow longitude, to which we fit a second-order
polynomial, yielding:
\begin{equation}
V_{flow}=163.4-2.854\lambda_{flow}+0.01364\lambda_{flow}^2,
\end{equation}
which applies for a longitude range of
$\lambda_{flow}=[72^{\circ},82^{\circ}]$.  Within this range the
relation agrees very well with a similar relation quoted by
\citet{djm12}.  The best-fit $\chi^2$ value shows
relatively little variation along this relation, making it difficult
to determine where along the relation is the true best fit.

     As we turn our attention from Maxwellian VDFs to bi-Maxwellians,
we actually use Equation~(3) to remove $V_{flow}$ as a free parameter
of our bi-Maxwellian fits.  Decreasing the number of free parameters
in this fashion is desirable given that a bi-Maxwellian VDF adds three
new fit parameters, which in turn makes the fits less robust.  The
term ``robust'' in this context relates to the ability of our fitting
routine to find its way to the same best fit regardless of our
initial guess for the best-fit parameters.  We conducted a number of
tests to confirm that the change from a Maxwellian to a bi-Maxwellian
does not affect the best-fit flow direction or magnitude, validating
our use of Equation~(3) in reducing the number of free parameters by
one.

     A bi-Maxwellian analysis is still impractical without
first doing a full grid search of the $(\lambda_{axis},b_{axis})$
parameter space to search for local minima in $\chi^2$.
With $(\lambda_{axis},b_{axis})$ fixed, and with $V_{flow}$ set by
Equation~(3), the remaining bi-Maxwellian fit parameters of interest
are the flow direction $(\lambda_{flow},b_{flow})$, $T_{\perp}$, and
$T_{\parallel}$.  The fits are now sufficiently robust for us to begin
the process of zeroing in on the best bi-Maxwellian fit to the {\em IBEX}
data.  By systematically varying the $(\lambda_{axis},b_{axis})$
orientation of the bi-Maxwellian, we produce a map of $\chi_{\nu}^2$
with $(\lambda_{axis},b_{axis})$.  The result is shown in Figure~3(a),
which shows a very clear $\chi^2$ minimum at
$(\lambda_{axis},b_{axis})=(57^{\circ},-2^{\circ})$.  A
bi-Maxwellian has $180^{\circ}$ symmetry, so the two minima in
Figure~3(a) are really the same minimum.  In this paper, we
will use the convention of quoting only the direction with the longitude
closest to $0^{\circ}$.  This map is actually a
product of two $(\lambda_{axis},b_{axis})$ grids, one with an angular
resolution of $30^{\circ}$ covering the full map, and one with a
higher angular resolution of $10^{\circ}$ surrounding the $\chi^2$
minimum region.
\begin{figure}[t]
\plotfiddle{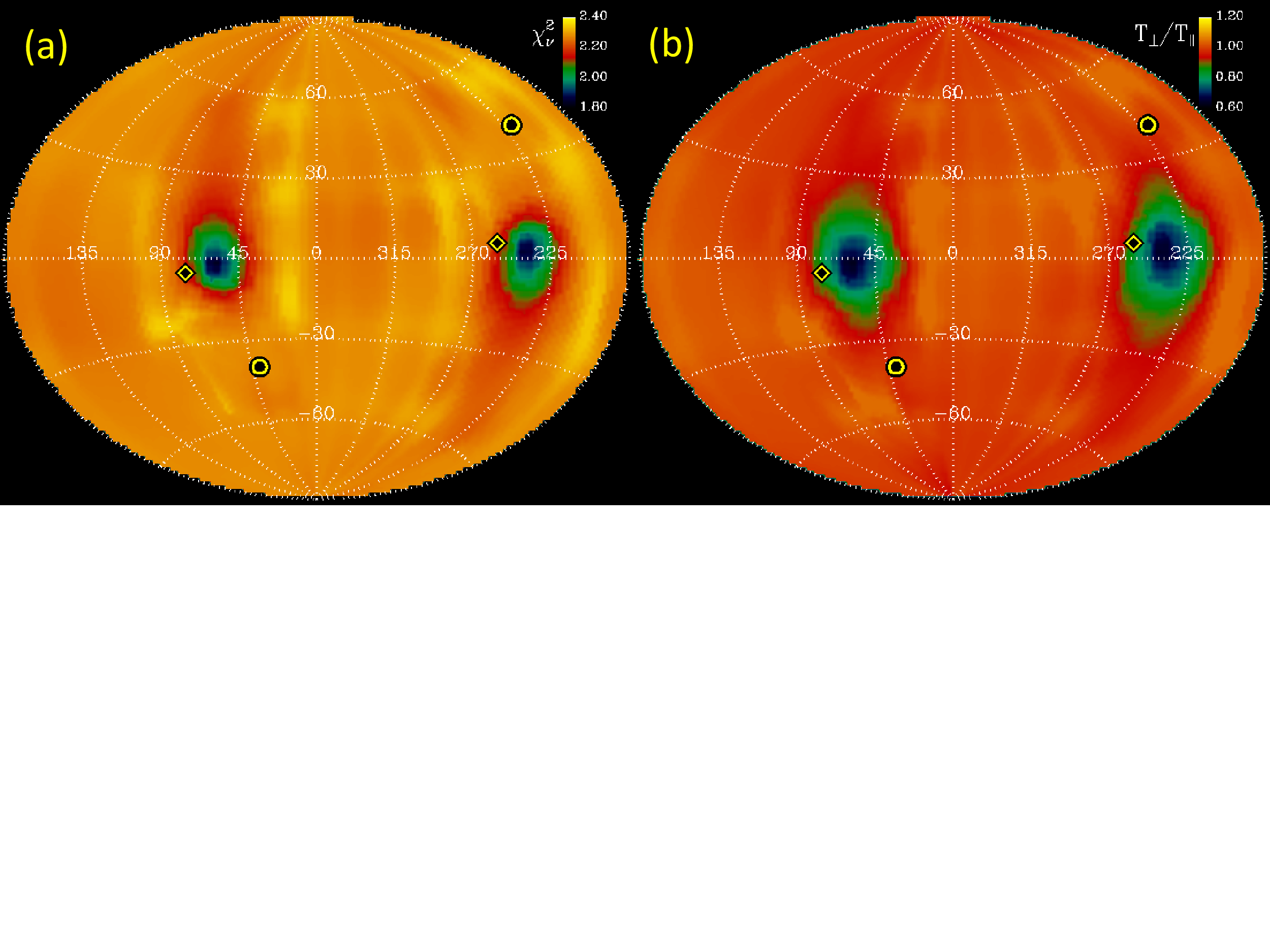}{2.6in}{0}{65}{65}{-235}{-165}
\caption{(a) For the {\em IBEX} data, a map of best-fit
  $\chi^2_{\nu}$ versus bi-Maxwellian orientation
  $(\lambda_{axis},b_{axis})$, showing a clear minimum at
  $(\lambda_{axis},b_{axis})=(57^{\circ},-2^{\circ})$, close to
  the $V_{flow}$ direction indicated by the diamond.
  (Due to the $180^{\circ}$ symmetry of the bi-Maxwellian, the minimum
  at $(237^{\circ},2^{\circ})$ is redundant.)  The circle indicates
  the $B_{ISM}$ direction suggested by the {\em IBEX} ribbon (Funsten
  et al.\ 2013).  (b) A map of best-fit $T_{\perp}/T_{\parallel}$ with
  $(\lambda_{axis},b_{axis})$, showing a clear $T_{\perp}/T_{\parallel}$
  minimum corresponding to the $\chi^2$ minimum in (a).}
\end{figure}

     Further support for the bi-Maxwellian solution at
$(\lambda_{axis},b_{axis})=(57^{\circ},-2^{\circ})$ is
provided by the map of best-fit $T_{\perp}/T_{\parallel}$ ratio shown
in Figure~3(b).  This map shows a clear minimum at exactly the same
location as the $\chi^2$ minimum.  This
supports the validity of the bi-Maxwellian solution, because if the He
VDF is truly bi-Maxwellian with $T_{\perp}/T_{\parallel}<1$, then in a
map like Figure~3(b) there should be a minimum in best-fit
$T_{\perp}/T_{\parallel}$ only where the correct
$(\lambda_{axis},b_{axis})$ is assumed.  If an incorrect bi-Maxwellian
axis direction is assumed, the best-fit $T_{\perp}/T_{\parallel}$
value will be closer to one.  It is only when the correct axis
orientation is assumed that the true low temperature ratio is
recovered.  Figure~3(b) is by itself suggestive of a preferred fit
with $T_{\perp}/T_{\parallel}<1$, at $(\lambda_{axis},b_{axis})=
(57^{\circ},-2^{\circ})$, regardless of whether there is a $\chi^2$
minimum at that location.  It is worth noting that Figure~3(b) shows
no evidence for a preferred fit anywhere with
$T_{\perp}/T_{\parallel}>1$.

\begin{figure}[t]
\plotfiddle{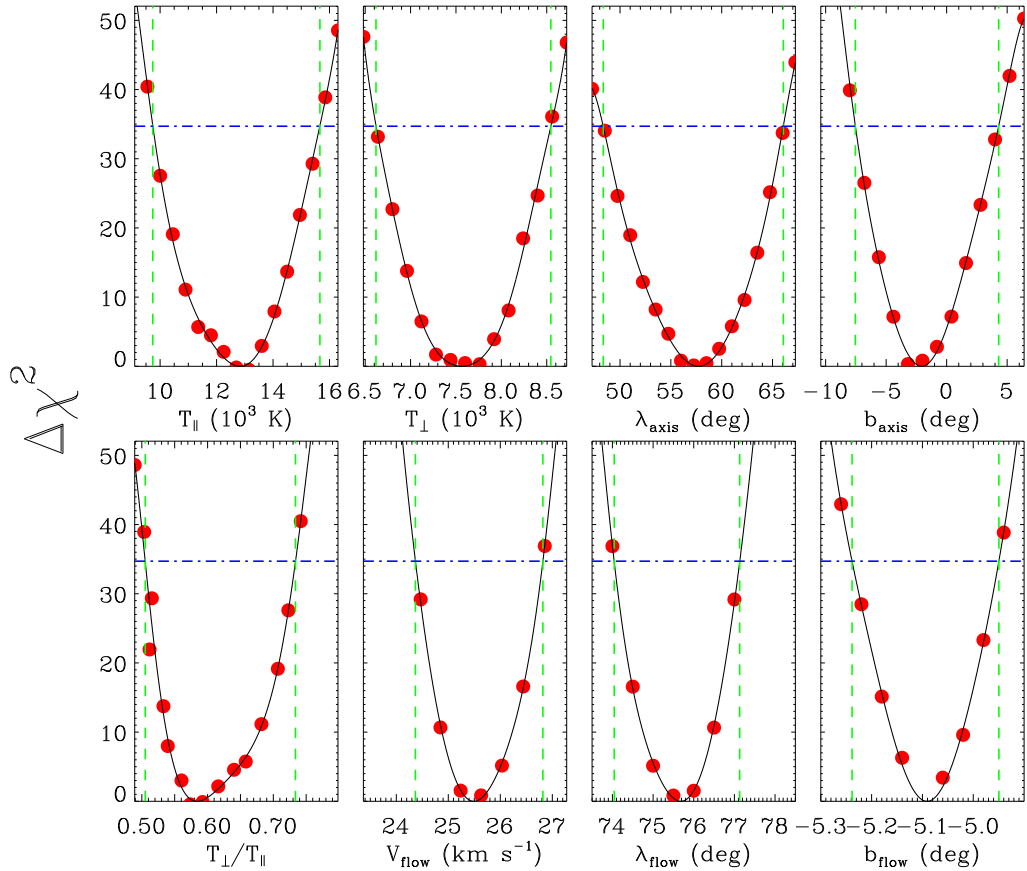}{4.4in}{90}{70}{70}{280}{-50}
\caption{An assessment of best-fit values and uncertainties for the
  bi-Maxwellian He flow parameters inferred from {\em IBEX} data.  In each
  panel, $\Delta\chi^2$ is plotted as a function of one of the
  parameters of interest.  Each point corresponds to a fit where that
  parameter is held constant, but the other parameters
  are allowed to vary.  Solid lines are high order polynomial fits
  to the $\Delta\chi^2$ values.  The horizontal dot-dashed
  line indicates our assumed $\Delta\chi^2$
  threshold, which is used to define the region of acceptable fit,
  shown explicitly as vertical dashed lines.}
\end{figure}
     With Figure~3 clearly identifying the location of the best-fit
bi-Maxwellian axis, we now conduct an error analysis for
the various bi-Maxwellian parameters of interest, which is shown in
Figure~4.  The temperature ratio, $T_{\perp}/T_{\parallel}$, is not an
independent parameter from the $T_{\perp}$ and $T_{\parallel}$
temperatures, but as a quantity of interest it is included in Figure~4
as well.  For each parameter, the analysis involves a sequence of fits
with that parameter held constant but the others allowed to vary, with
the assumed value of the parameter then stepped across the $\chi^2$
minimum value to see at what point the increase in $\chi^2$,
$\Delta\chi^2=\chi^2-\chi^2_{min}$, becomes too large to be considered
acceptable.  The acceptable $\Delta\chi^2$ value depends on the number
of free parameters of the fit, which include the 7 bi-Maxwellian
parameters of interest, plus 4 background and 4 normalization factors
for the 4 years of data considered (see above).  Based on these 15
free parameters, we assume a $3\sigma$ confidence contour
corresponding to $\Delta\chi^2=34.7$, based on relation 26.4.14 of
\citet{ma65}.  The error bars displayed in Figure~4 are
based on this threshold.

     The final bi-Maxwellian parameters and uncertainties are listed
in Table~1, where they are compared with recent Maxwellian measurements
from both {\em Ulysses} \citep{bew15} and {\em IBEX} \citep{mb15}.
The flow vector
($V_{flow}$, $\lambda_{flow}$, $b_{flow}$) is in very good agreement
across all these analyses.  Changing from a Maxwellian VDF to a
bi-Maxwellian clearly has no effect on the flow vector.  It
is instructive to compare our {\em IBEX} He flow parameter
uncertainties with those that we previously measured for {\em Ulysses}
\citep{bew15}, because the uncertainties have been computed in a
consistent fashion, with consistent $3\sigma$ $\Delta\chi^2$ thresholds.
The $V_{flow}$ and $\lambda_{flow}$ uncertainties
are significantly lower for {\em Ulysses} than for {\em IBEX}, despite
the significantly higher signal-to-noise (S/N) ratio of the {\em IBEX}
data.  The causes of the large {\em IBEX} uncertainties are the
parameter degeneracies alluded to above, and explicitly provided in
Equation~(3).  Unlike {\em IBEX}, which essentially makes a single He
beam map each year from the same location, the {\em Ulysses} mission
observed the He beam at many different locations in the inner
heliosphere.  As long as these data are considered collectively, the
{\em Ulysses} measurements are free from the parameter degeneracy
problem \citep{bew15}, leading to smaller parameter
uncertainties despite the lower S/N.

     In the bi-Maxwellian fit, the $T_{\perp}=7580\pm 960$~K value is
very similar to the temperature inferred in the Maxwellian
analyses.  It is $T_{\parallel}$ that changes greatly, increasing
to $T_{\parallel}=12,700\pm 2960$~K.  Perhaps this might provide
an explanation for why certain analyses of {\em IBEX} data have ended
up preferring higher temperatures, such as that of \citet{em15},
who quote $T=8710^{+440}_{-680}$~K.  If
there is indeed $T$ anisotropy in the He VDF, then different types
of analyses that assume a simple Maxwellian VDF could be sensitive
to $T_{\perp}$ and $T_{\parallel}$ to different degrees.

     Although of less interest than the He flow parameters, for
completeness we now list the best-fit backgrounds and normalization
factors for the four considered years of {\em IBEX} data.  For the
former, our best-fit background count rates are 0.0145, 0.0235,
0.0208, and 0.0077 s$^{-1}$ for 2011--2014, respectively.  Without a
precise calibration, the
normalization factors have arbitrary units, so we divide them by their
mean and report values of 1.346, 1.229, 0.675, and 0.750 for
2011--2014, respectively.  The cause of the factor of two drop after
2012 is a reduction in the post-acceleration voltage at that time,
which changed the instrument sensitivity \citep[e.g.,][]{mb15}.

\begin{figure}[t]
\plotfiddle{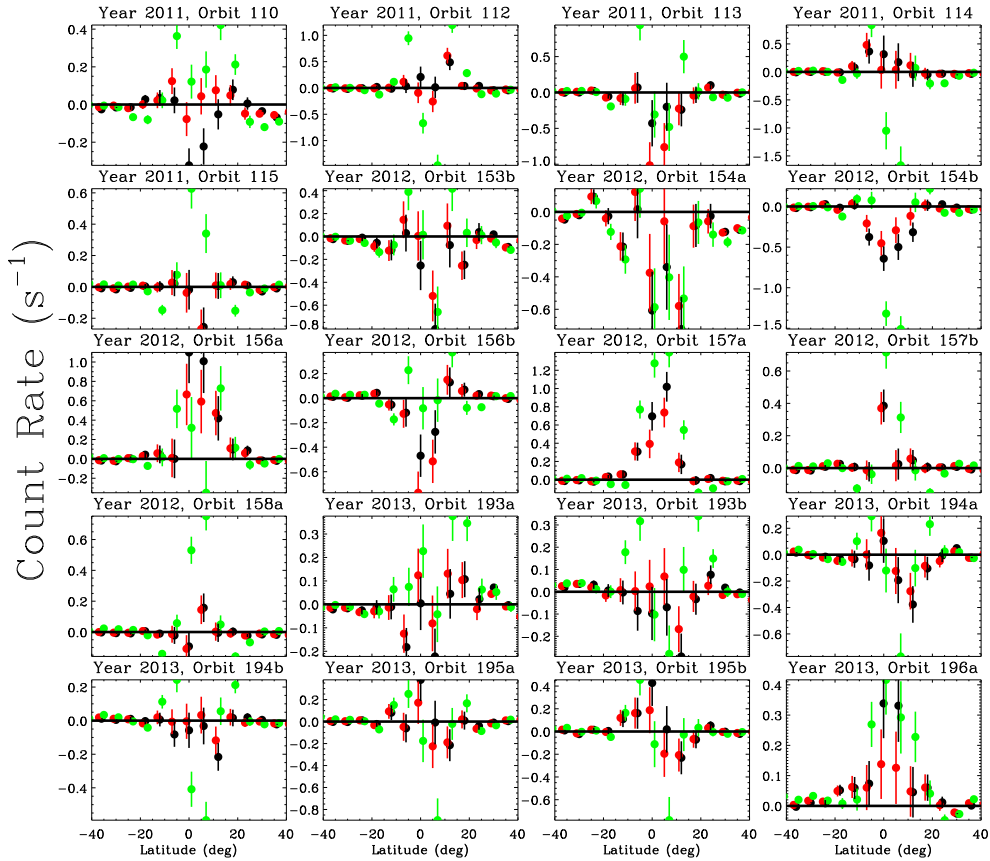}{4.5in}{90}{70}{70}{280}{-50}
\caption{The residuals of three fits to IBEX data, with the black points
  the best fit assuming a Maxwellian VDF, the red points the best
  bi-Maxwellian fit, and the green points the best kappa
  distribution fit with $\kappa=3$ (see Section~3).}
\end{figure}
\setcounter{figure}{4}
\begin{figure}[t]
\plotfiddle{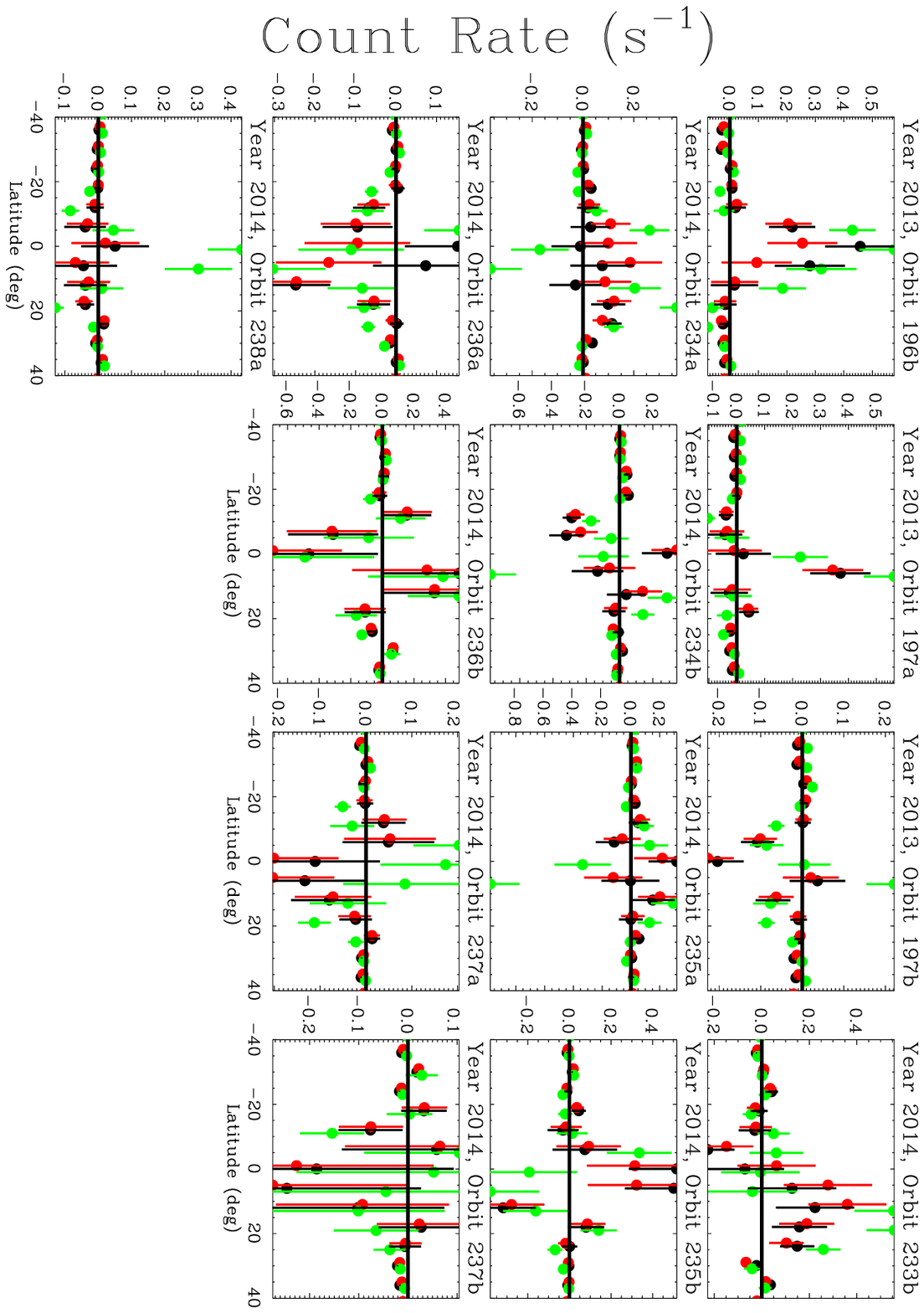}{3.7in}{90}{70}{70}{280}{-110}
\caption{(continued)}
\end{figure}
     The best bi-Maxwellian fit to the {\em IBEX} data has
$\chi^2_{\nu}=1.83$, a significant improvement over the
best Maxwellian fit with $\chi^2_{\nu}=2.28$.  In order to
try to identify where the bi-Maxwellian model is fitting the data
better than the Maxwellian one, Figure~5 compares residuals
of the fits for all 33 orbits considered in 2011--2014.
It is difficult to distinguish between the fits in a figure
like Figure~1, which is why we here show only the residuals.

     Unfortunately, Figure~5 fails to provide a clear indication
as to what aspects of the IBEX measurements are being reproduced
better by the bi-Maxwellian model than by the Maxwellian one.  The
most fundamental problem is that the residual pattern is
not precisely repeated from one year to the next.  For example,
the overestimate of flux apparent in the 2011 orbits 156a and 156b
is not necessarily seen in analogous orbits in other years
that correspond to the same position in Earth's orbit around
the Sun (e.g., orbits 196b and 197b in 2012).  Some of this
irreproducibility is due to the good time intervals in one
orbit being significantly different from the good time intervals
in the corresponding orbit in other years, meaning that the
observations may be from different characteristic locations in
{\em IBEX}'s orbit around the Earth, leading to different
systematic effects on the data.  Nevertheless,
there are some orbits like 196a and 196b where the bi-Maxwellian
fit is clearly doing significantly better than the Maxwellian fit.

\begin{figure}[t]
\plotfiddle{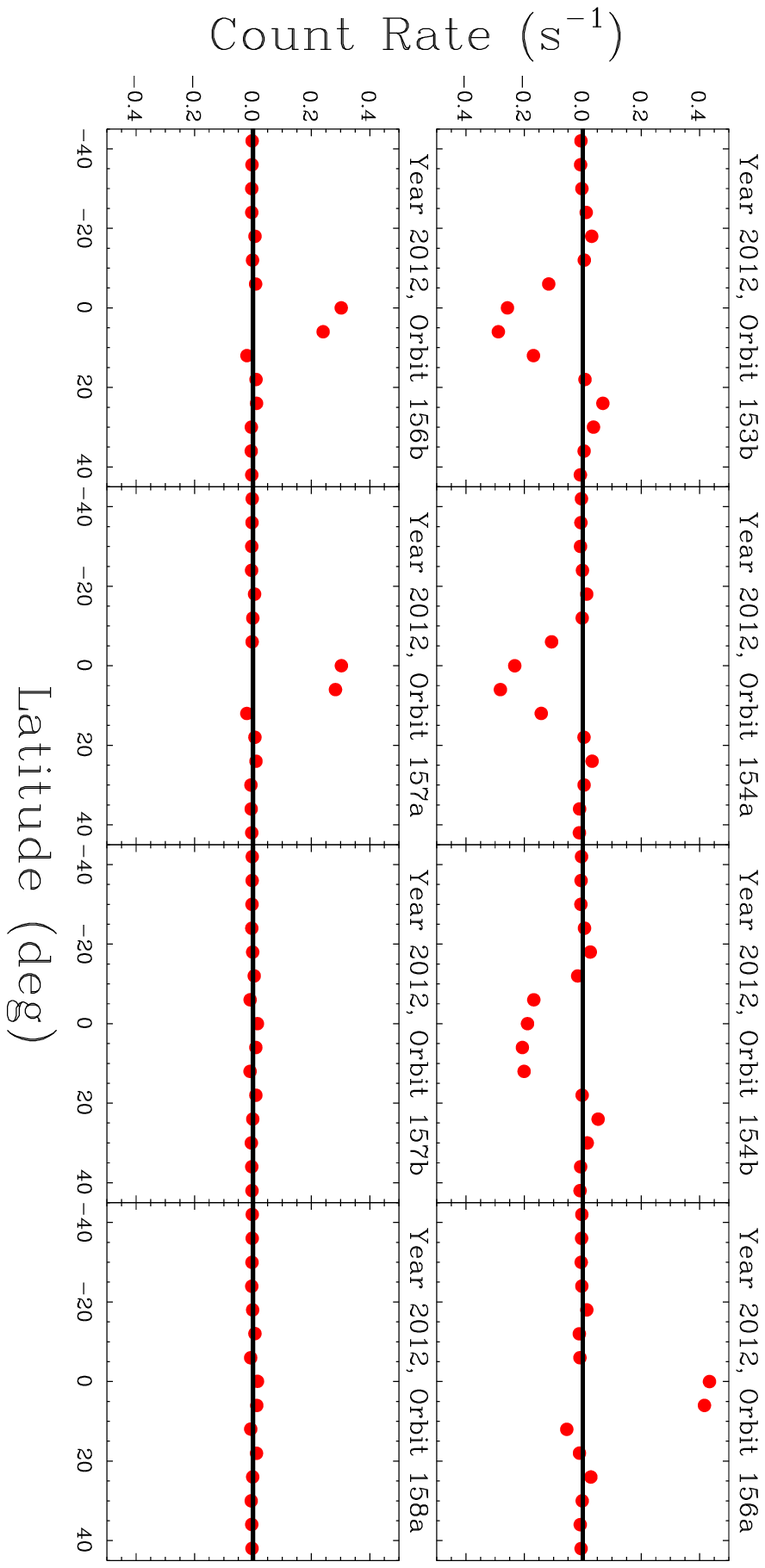}{3.2in}{90}{90}{90}{370}{-260}
\caption{Using the 2012 orbits as an example, we plot the count rate
  differences predicted by the best-fit bi-Maxwellian and
  single-Maxwellian fits.  Specifically, the single-Maxwellian
  count rates are subtracted from the bi-Maxwellian rates.  The
  bi-Maxwellian fit predicts lower rates at early orbits, higher
  rates for the middle orbits, and similar rates for the last
  orbits.  All deviations are between 1.5\% and 4\%.
  A similar pattern is seen in other years.}
\end{figure}
     Figure~6 provides a clearer illustration of how the
best-fit bi-Maxwellian count rates compare to the best-fit single
Maxwellian count rates, using year 2012 as an example, as in Figure~1.
The bi-Maxwellian has lower rates at early orbits in the observing
season, followed by higher rates in the middle orbits.  There is
little difference in the fits for the last orbits.  This basic
pattern is reproduced each year.

\begin{figure}[t]
\plotfiddle{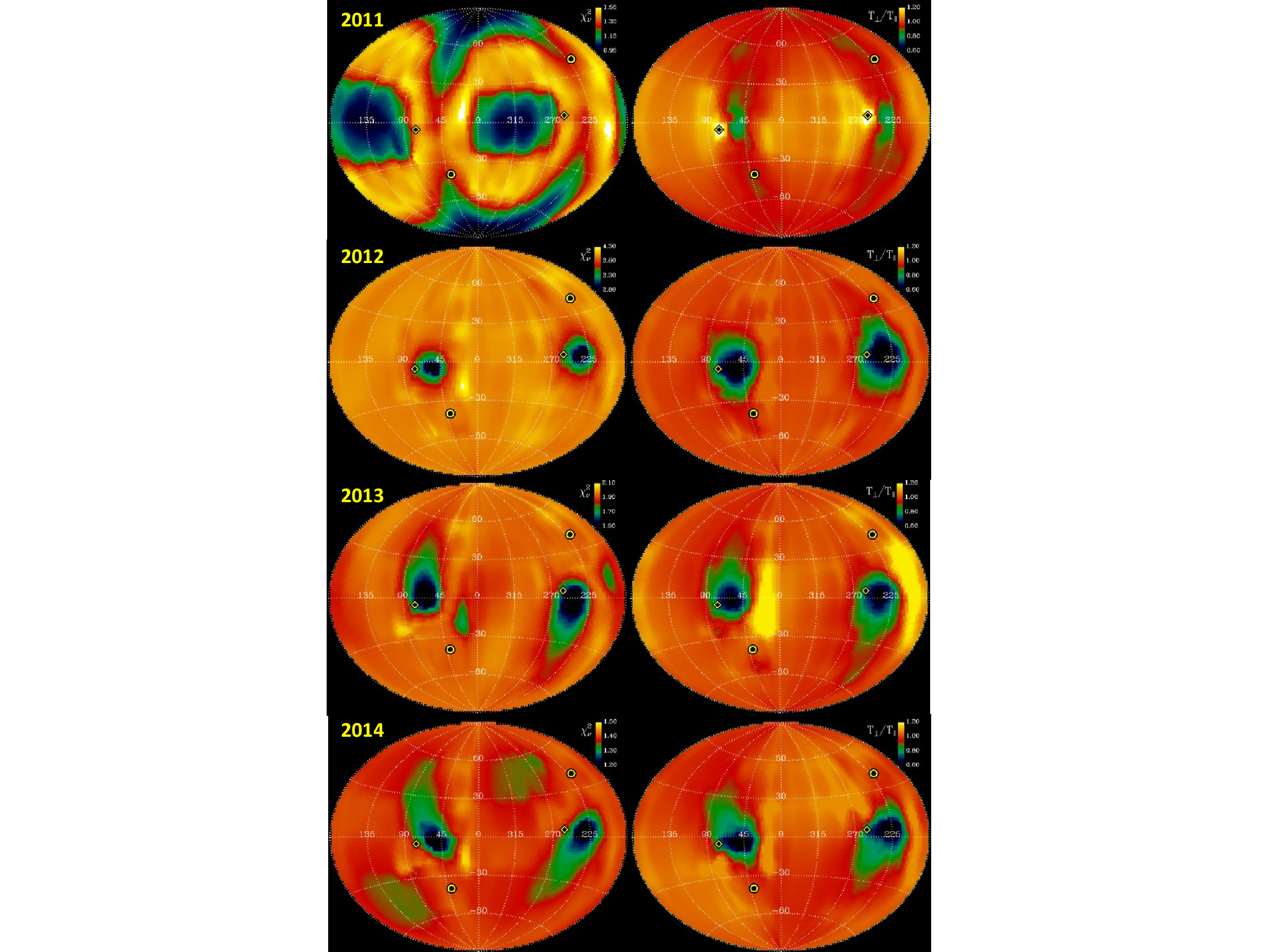}{6.6in}{0}{85}{85}{-300}{0}
\caption{Analogous to Figure~3, but for the four considered years
  of IBEX data analyzed separately.  All four years show the same
  $T_{\perp}/T_{\parallel}$ minimum region seen in Figure~3, although
  in 2011 this minimum is much less pronounced and there is no
  $\chi^2_{\nu}$ minimum at that location.}
\end{figure}
     In order to further explore how and why the analysis is
ending up at the bi-Maxwellian solution indicated in Figure~3, we
conduct the analysis for the four years (2011--2014) separately,
to see if all four years independently suggest the
same best fit.  The results are shown in Figure~7.  As in
Figure~3, we use a $30^{\circ}$ resolution grid for the full
map, but a higher $10^{\circ}$ resolution grid around the
$T_{\perp}/T_{\parallel}$ minimum region of interest, with
the merging and interpolation process involved in combining
the two grids introducing some artifacts.  The
$T_{\perp}/T_{\parallel}$ maps are shown on the same color scale for
ease of comparison.  This is not possible for the $\chi^2_{\nu}$
maps, since $\chi^2_{\nu}$ is quite different for the different
years.

     All four maps in Figure~7 show a $T_{\perp}/T_{\parallel}$
minimum region near $(\lambda_{axis},b_{axis})=(57^{\circ},-2^{\circ})$,
similar to that in Figure~3.  Thus, we conclude that all four
years are independently supporting the bi-Maxwellian fit suggested
by the collective analysis in Figure~3.  By far the biggest
discrepancies with the Figure~3 maps are seen in 2011, where the
$T_{\perp}/T_{\parallel}$ minimum is much less pronounced and
there is no longer any $\chi^2_{\nu}$ minimum at all at that
location.

     It is necessary to discuss the possibility that the apparent
success of the bi-Maxwellian fit might be due to systematic errors in
the data or artifacts of an imprecise analysis, rather than being an
indication of asymmetries in the actual He VDF.  The signature of the
preferred bi-Maxwellian fit in Figure~3 seems very strong, but the
difficulty in clearly identifying in the data why this particular fit
works significantly better than the Maxwellian model is worrisome.
There is also some reason for concern that the bi-Maxwellian fit
results could be indicative of inconsistencies in two distinct
temperature constraints provided by the {\em IBEX} data.  Using
Figure~1 as an example, the first constraint is the width of the
latitudinal scans seen in the individual panels of the figure, while
the second constraint is the width of the peak-flux-versus-orbit
behavior seen when considering all panels together.  Increasing a
Maxwellian temperature should broaden the distribution in both senses,
but in a bi-Maxwellian paradigm the two distinct constraints may have
different sensitivities to $T_{\perp}$ and $T_{\parallel}$.  It is
perhaps worrisome that the inferred $(\lambda_{axis},b_{axis})$
direction is very close to the ecliptic where {\em IBEX} is observing,
and is also close to the flow direction, $(\lambda_{flow},b_{flow})$.
With this geometry, the $T_{\perp}$ parameter will primarily
define the latitudinal scan width.  Both $T_{\perp}$ and
$T_{\parallel}$ will influence the peak-flux-versus-orbit
width, the former through the distribution of angular particle
directions in the ecliptic, and the latter through the difference in
degree of gravitational deflection for particles of different energies
directed along the central flow direction.

     We have made an effort to change our analysis in various
ways to see how sensitive the results in Figure~3 are to various
assumptions that have to be made.  We increased the number of
orbits considered in the fit, and also considered a broader
latitudinal range.  We considered an analysis in which the
secondary He component contribution was allowed to be a free
parameter of the fit, and an analysis that makes no correction
for secondaries at all.  None of these experiments yielded any
significant change to the maps in Figure~3.  They still suggest
a preferred fit with
$(\lambda_{axis},b_{axis})=(57^{\circ},-2^{\circ})$,
and $T_{\perp}/T_{\parallel}<1$.

     Our analysis follows the practice of most past IBEX analyses
in assuming no energy dependence for the He detection efficiency.
However, we now briefly explore how our results might
be affected by an energy-dependent sensitivity function, following
the example of \citet{ps18}.  \citet{ps18} assume a sensitivity
function, $S_{PV}$, that is linearly dependent on particle velocity,
\begin{equation}
S_{PV}\propto 1+b_p(v-v_0),
\end{equation}
where $v_0=78$ km~s$^{-1}$ is an arbitrary reference velocity, which
is a typical speed for He particles when they reach IBEX (see
Figure~2).  We assume a value of $b_p=-0.05$ km$^{-1}$~s, a typical
value found by \citet{ps18} when they allow $b_p$ to vary in fits
to data, implying a higher
sensitivity to slower He particles.  We then repeat the bi-Maxwellian
fit, leading to the new maps in Figure~8, which can be compared
with our baseline results in Figure~3.

\begin{figure}[t]
\plotfiddle{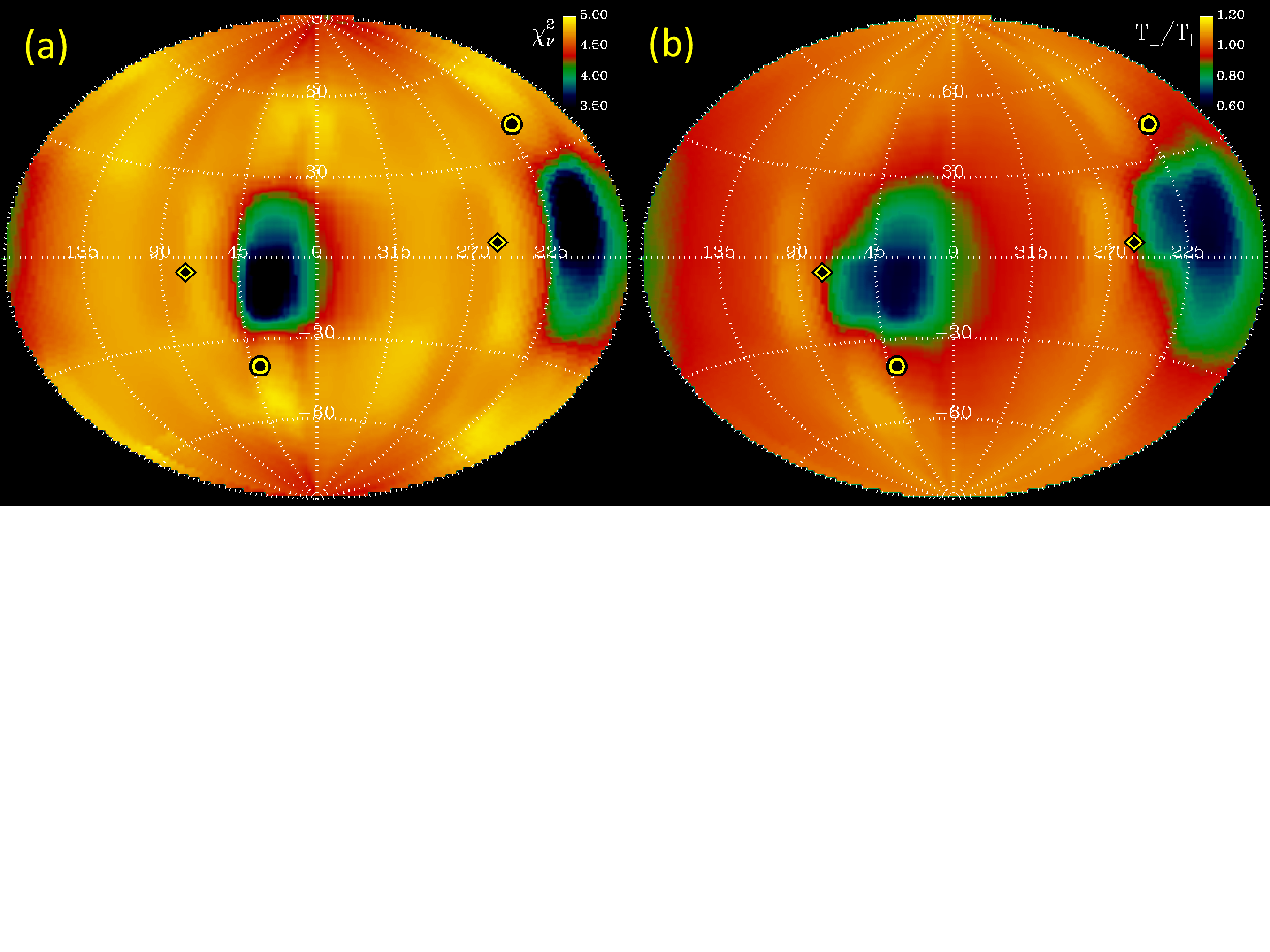}{2.6in}{0}{65}{65}{-235}{-165}
\caption{Analogous to Figure~3, but for a
  model with an assumed energy dependence for the He detection
  efficiency.  A clear best fit bi-Maxwellian solution
  with $T_{\perp}/T_{\parallel}<1$ persists, but its location
  has shifted in longitude compared to Figure~3.}
\end{figure}
     With this assumed energy dependence for the
detector sensitivity we still find a best-fit bi-Maxwellian with
$T_{\perp}/T_{\parallel}<1$, where $\chi^2$ is minimized.
However, the longitude of the minimum
region, $\lambda_{axis}$, is shifted.  It is important to note that
none of the other bi-Maxwellian fit parameters listed in Table~1 are
significantly affected at all.  This is consistent with the results
of \citet{ps18}, who found no significant change in best-fit He flow
direction or magnitude when the detector sensitivity was allowed
to depend on particle velocity, as in equation~(4).  However, in
our bi-Maxwellian analysis, we find that $\lambda_{axis}$ decreases
from $\lambda_{axis}=57.2^{\circ}\pm 8.9^{\circ}$ to
$\lambda_{axis}\approx 30^{\circ}$.
Although this exercise demonstrates that it is possible that
$\lambda_{axis}$ in a bi-Maxwellian analysis could be affected by
an energy sensitivity in particle detection, we note that the
quality of the best fit within Figure~8 ($\chi^2_{\nu}=3.35$)
is significantly worse than the best fit in Figure~3 assuming no
energy sensitivity ($\chi^2_{\nu}=1.83$).  Further work
allowing a variable $b_p$ would be required to demonstrate that
an assumed energy
dependence of instrument sensitivity is statistically justified
in the context of a bi-Maxwellian fit, which is outside the
scope of our current analysis.

     The one experiment we tried that showed potential for
yielding truly fundamental changes to our bi-Maxwellian fit
results was to vary the assumed photoionization rates.
Our first experiment was to remove the loss rate correction entirely.
This not only did not change the basic character of Figure~3,
but it also greatly worsened the quality-of-fit to $\chi^2_{\nu}\sim 5$.
However, when we go the opposite direction and arbitrarily
increase the photoionization loss rates by 50\%, we have more promising
results, which are shown in Figure~9.  The appearance of the maps
in Figure~9 is curiously similar to those for
2011 in Figure~7, suggesting similar systematic uncertainties may
be coming into play here.  Could this mean that the photoionization
rate assumed for 2011 is significantly too high?  In Figure~9,
with the 50\% increase in photoionization, the $\chi^2$ minimum near
$(\lambda_{axis},b_{axis})=(57^{\circ},-2^{\circ})$
disappears entirely.  The $T_{\perp}/T_{\parallel}$ minimum at that
location fades as well, although it is still visible.  Furthermore,
the $\chi^2_{\nu}$ values found for these fits do not indicate any
dramatic degradation in fit quality like we see when photoionization
rates are arbitrarily decreased instead of increased.
\begin{figure}[t]
\plotfiddle{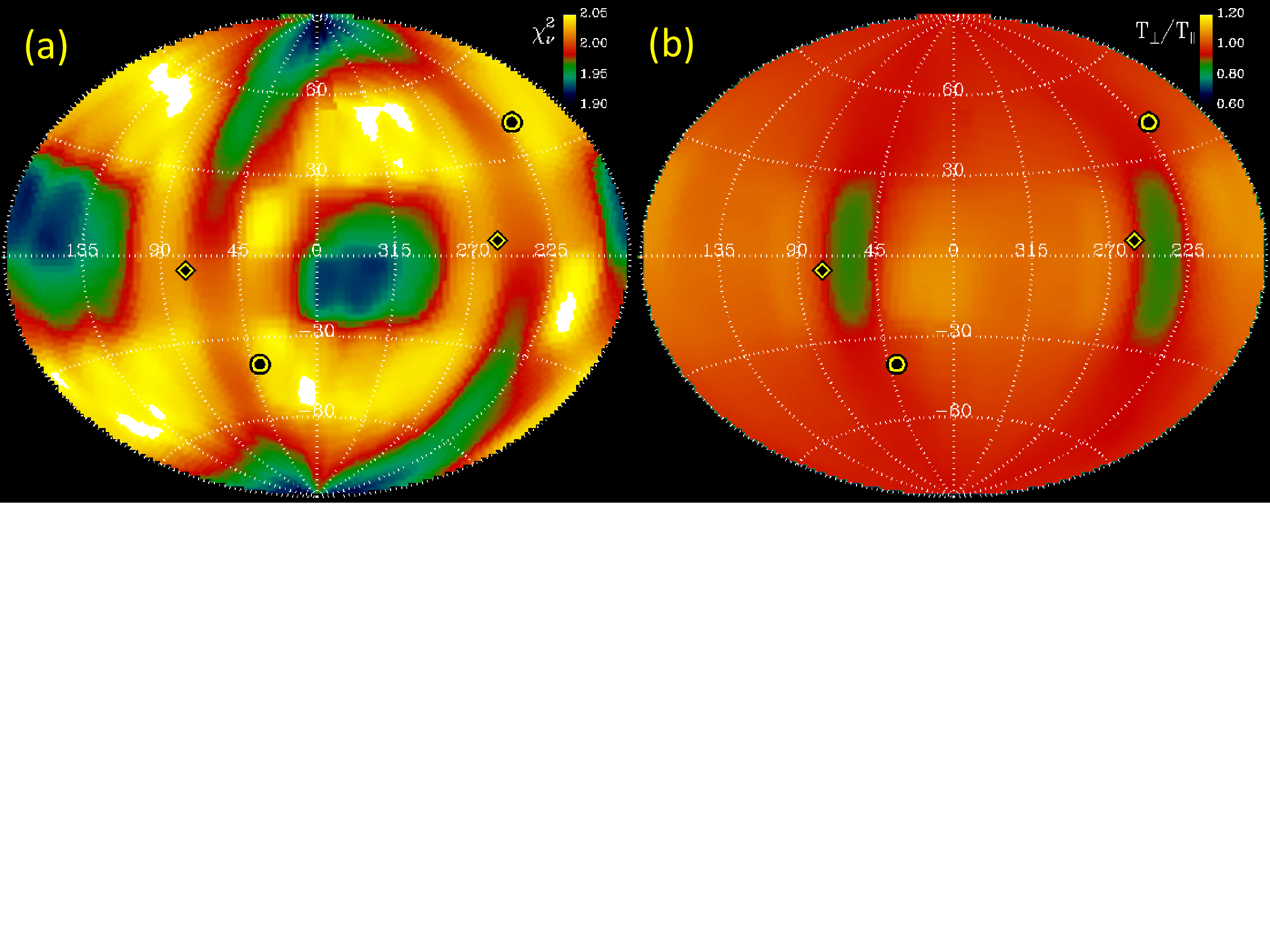}{2.6in}{0}{65}{65}{-235}{-165}
\caption{Analogous to Figure~3, but for a
  model with photoionization rates arbitrarily increased by 50\%.
  The $\chi^2$ minimum at $(\lambda_{axis},b_{axis})=
  (57^{\circ},-2^{\circ})$ has disappeared, and the
  $T_{\perp}/T_{\parallel}$ minimum at that location has
  faded, although it is still visible.}
\end{figure}

     Thus, our best bi-Maxwellian fit with $T_{\perp}/T_{\parallel}<1$
and $(\lambda_{axis},b_{axis})=(57^{\circ},-2^{\circ})$
could in principle be an artifact of underestimated photoionization
loss rates, but it is questionable whether the rates could really be
systematically that far off.  We use a simplified treatment of
the loss rates here, where a single constant rate is used for all particles
in an observing season, with no attempt to account for rate variation
during the He atoms' slow journey to 1~AU.  \citet{jms15}
studied the effects of loss rate assumptions in great detail,
including the effects of secondary loss processes such as electron
impact ionization and charge exchange near the Sun, and including the
effects of latitudinal variation.  Their analysis implies that our
assumption of constant loss rates could lead to loss estimates
being of order 7\% off for a given observing season.
However, this is still much lower than the $\sim 50$\% error needed
to affect our analysis.  Furthermore, such errors should not be
consistent from year to year, so the persistence of the
$T_{\perp}/T_{\parallel}$ minimum region for all four years in
Figure~7 implies that the effects of loss rate variability are not
significant factors.  Only a latitude dependence
issue would likely lead to a systematic error in the same
direction every year, but the analysis of \citet{jms15} implies
that errors introduced by that effect are less than 1\%.


     Our preferred interpretation remains that the bi-Maxwellian
results in Figure~3 are truly indicative of VDF asymmetries in the He
flow that exist far from the Sun, but there is clearly value in
looking for independent support for these asymmetries elsewhere.  With
that in mind, we turn our attention to {\em Ulysses}.



\subsection{{\em Ulysses} Data Analysis}

     The GAS instrument on the {\em Ulysses} spacecraft \citep{mw92}
conducted episodic observations of the ISN He flow
between the 1990 launch date and the end of the mission in 2007.  The
primary observations were made from {\em Ulysses}'s operational orbit
roughly perpendicular to the ecliptic plane, with an aphelion near
5~AU and a perihelion near 1~AU.  The He observations could only be
made during the fast latitude scans near perihelion, with the
spacecraft moving fast enough for He particles to be entering the GAS
detector with sufficient energy for detection.  Thus, the observations
are limited to three periods from 1994--1996, 2000--2002, and
2006--2007.

     When the He flow was observable, the GAS instrument could make
a scan across the He beam in a $2-3$ day period, so the {\em Ulysses}
database consists of hundreds of separate He beam maps from many
different places in {\em Ulysses}'s orbit.  Based on a set of 238 of
these maps, we have previously inferred He flow properties assuming a
Maxwellian VDF \citep{bew15}.  Our analysis procedures here are
essentially identical to this previous work, only with the Maxwellian
VDF replaced with a bi-Maxwellian.  We refer the reader to \citet{bew15}
for details.  Before fitting the data, we subtract the
secondary He contribution reported by \citet{bew17}.

\begin{figure}[t]
\plotfiddle{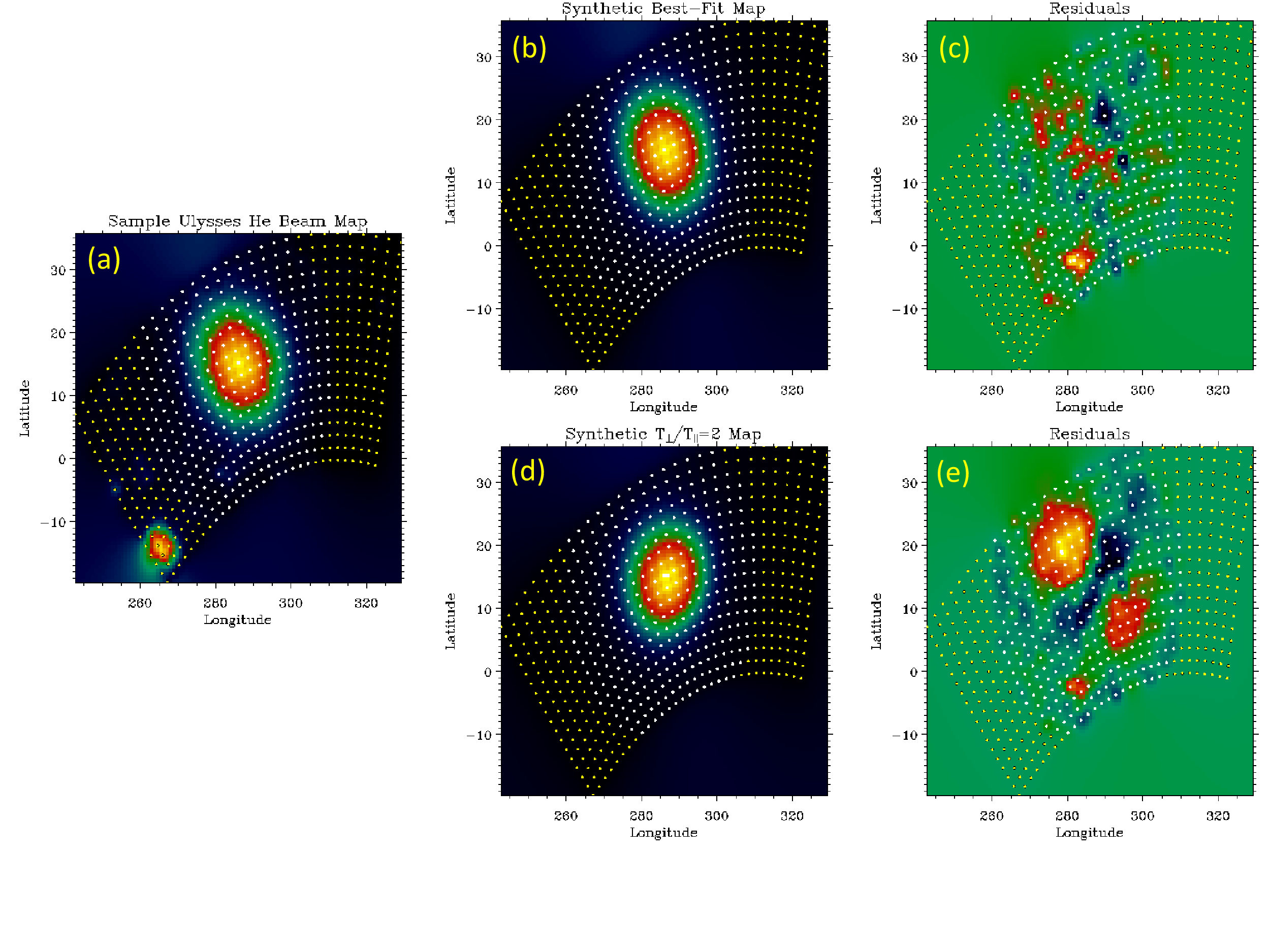}{4.0in}{0}{65}{65}{-235}{-50}
\caption{(a) A sample {\em Ulysses} He beam map, from 2001~February~4,
  with the dots indicating scan positions and the white dots explicitly
  indicating the points used in fitting the data.  (b) A synthetic He
  beam computed assuming the best-fit He flow vector.  (c) Residuals
  of the synthetic beam from (b) subtracted from the data in (a).
  (d) A synthetic He beam computed assuming the best-fit He flow vector,
  but with the introduction of a $T_{\perp}/T_{\parallel}=2$ temperature
  anisotropy. (e) Residuals of
  the synthetic beam from (d) subtracted from the data in (a).}
\end{figure}
     Figure~10(a) shows an example of a {\em Ulysses} He beam map, from
2001~February~4, and the best fit to this data from \citet{bew15} in
Figure~10(b), with the residuals in Figure~10(c).  In order to
illustrate the effects of a bi-Maxwellian distribution with a large
temperature anisotropy, in Figure~10(d) we simulate a beam
map assuming the best-fit He flow vector, but with the
temperature changed so that $T_{\perp}/T_{\parallel}=2$.  Specifically,
we replaced the best-fit single Maxwellian $T=7260$~K with
$T_{\perp}=7500$~K and $T_{\parallel}=3750$~K.  For this
purely illustrative test, we used the $B_{ISM}$ direction suggested
by the {\em IBEX} ribbon as the bi-Maxwellian axis,
$(\lambda_{axis},b_{axis})=(39.2^{\circ},-39.9^{\circ})$ \citep{hof13}.
The residuals in Figure~10(e) are naturally much worse than those
in Figure~10(c).

\begin{figure}[t]
\plotfiddle{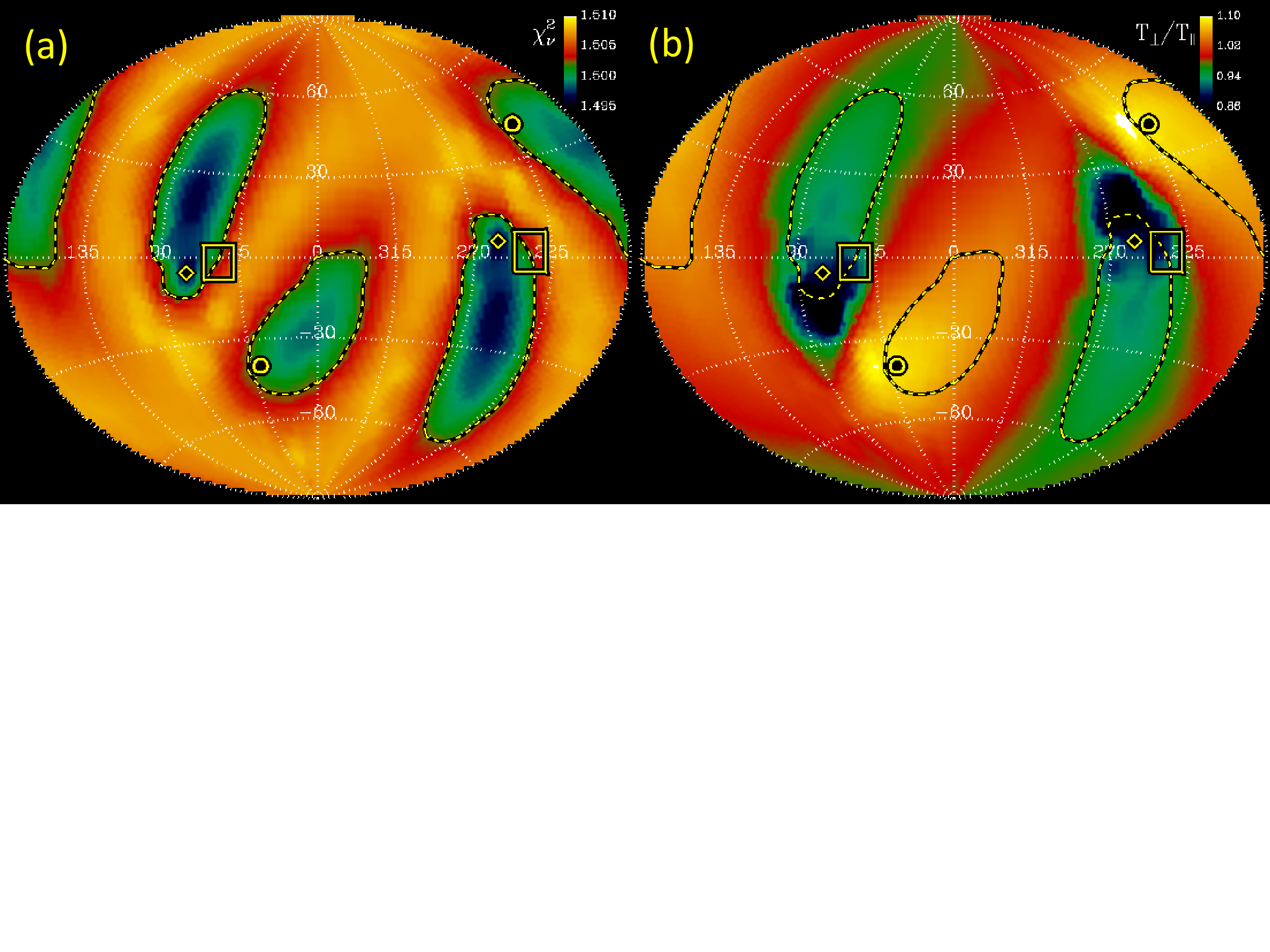}{2.6in}{0}{65}{65}{-235}{-165}
\caption{Analogous to Figure~3, but for a bi-Maxwellian fit to
  {\em Ulysses} data instead of {\em IBEX}.  There are two broad,
  distinct regions providing statistically acceptable fits, as
  defined by the $\chi^2_{\nu}$ threshold shown as a dashed line,
  one at $(\lambda_{axis},b_{axis})\approx
  (60^{\circ},15^{\circ})$ with $T_{\perp}/T_{\parallel}<1$, and
  one at $(\lambda_{axis},b_{axis})\approx (10^{\circ},-30^{\circ})$
  with $T_{\perp}/T_{\parallel}>1$.  The box is the best-fit
  $(\lambda_{axis},b_{axis})$ region from {\em IBEX} (see Table~1),
  which overlaps the $T_{\perp}/T_{\parallel}<1$ {\em Ulysses}
  region, indicating a degree of consistency between the
  {\em IBEX} and {\em Ulysses} bi-Maxwellian analyses.}
\end{figure}
     As was the case in the {\em IBEX} analysis, it is necessary to
do a full search through the $(\lambda_{axis},b_{axis})$ parameter
space to look for $\chi^2$ minima.  The resulting maps of
$\chi^2_{\nu}$ and best-fit $T_{\perp}/T_{\parallel}$ are shown in
Figure~11.  Unlike {\em IBEX}, the {\em Ulysses} map in Figure~11 shows
two distinct $\chi^2$ minima, one at $(\lambda_{axis},b_{axis})\approx
(60^{\circ},15^{\circ})$ with $T_{\perp}/T_{\parallel}<1$,
and one at $(\lambda_{axis},b_{axis})\approx (10^{\circ},-30^{\circ})$
with $T_{\perp}/T_{\parallel}>1$.
At the $\chi^2_{\nu}$ minima the best-fit
temperature values are:  $T_{\perp}=6930$~K, $T_{\parallel}=7440$~K, and 
$T_{\perp}/T_{\parallel}=0.93$ at $(\lambda_{axis},b_{axis})\approx
(60^{\circ},15^{\circ})$; and $T_{\perp}=7280$~K, $T_{\parallel}=6760$~K, and 
$T_{\perp}/T_{\parallel}=1.08$ at $(\lambda_{axis},b_{axis})\approx
(10^{\circ},-30^{\circ})$.
Using the same $3\sigma$ thresholds used
in the {\em IBEX} analysis in Figure~4, we compute a
$\chi^2_{\nu}=1.5025$ threshold for acceptable fits, which is
indicated as a dashed line in Figure~11.  Both $\chi^2$ minima regions
are statistically acceptable fits to the data, and the regions around
each minimum where the fits are acceptable are quite large.  This is
indicative of the low S/N of the {\em Ulysses} data, compared to IBEX.

     By itself, the {\em Ulysses} analysis
provides only marginal evidence that a bi-Maxwellian assumption
yields an improvement in quality of fit relative to a Maxwellian
assumption, with two rather broad $(\lambda_{axis},b_{axis})$ regions
where the fits are acceptable.  The best-fit bi-Maxwellian fit has
$\chi^2_{\nu}=1.497$, compared with $\chi^2_{\nu}=1.524$ for the
best Maxwellian fit \citep{bew15}.  This improvement is much
less than that seen for {\em IBEX} (see Section~2.1).
The temperature anisotropy is not quite as
large as we find for {\em IBEX}, but perhaps this is also a
manifestation of the lower statistical significance of the
improvement in fit quality provided by the bi-Maxwellian assumption.

     The most important aspect of Figure~11 to note is that the
$\chi^2$ minimum region of {\em Ulysses} with $T_{\perp}/T_{\parallel}<1$ 
does overlap with the bi-Maxwellian fit of {\em IBEX}, which also has
$T_{\perp}/T_{\parallel}<1$.  Thus, we conclude that the {\em Ulysses}
data do provide independent support for the reality of the best-fit
solution to the {\em IBEX} data in Figure~3.  Furthermore, the
{\em Ulysses} analysis is insensitive to assumptions about
photoionization loss rates \citep{bew15}, so the potential
dependence of the {\em IBEX} bi-Maxwellian results on the assumed
loss rates (see Figure~9) does not exist for {\em Ulysses}.

\subsection{Interpretation of the Bi-Maxwellian Results}

     We have found evidence that the ISN neutral He flow through
the inner heliosphere exhibits a non-Maxwellian character, with
the {\em IBEX} data being fit better with a bi-Maxwellian VDF
instead of a Maxwellian, with an orientation of
$(\lambda_{axis},b_{axis})=(57.2^{\circ}\pm 8.9^{\circ},
-1.6^{\circ}\pm 5.9^{\circ})$ and a temperature anisotropy of
$T_{\perp}/T_{\parallel}=0.62\pm 0.11$.
As noted at the beginning of Section~2, a bi-Maxwellian VDF model
makes the most physical sense if oriented about the magnetic field.
Thus, the case for the He flow being truly bi-Maxwellian would be
particularly strong if the best-fit ($\lambda_{axis}$,$b_{axis}$)
orientation had ended up close to the $B_{ISM}$ direction.

     The {\em IBEX} ribbon observations imply a $B_{ISM}$ direction
at or near $(39.2^{\circ},-39.9^{\circ})$ \citep{hof13}, a
direction explicitly indicated in both Figures~3 and 11.  However,
this direction is $41.6^{\circ}$ from the best-fit
$(\lambda_{axis},b_{axis})$ direction.
The {\em Ulysses} results in Figure~11 do show an acceptable
fit with $(\lambda_{axis},b_{axis})$ near the $B_{ISM}$ direction,
with $T_{\perp}/T_{\parallel}>1$, but {\em IBEX} provides no
support for this solution.  Thus, the actual He VDF implied by
our results may not be truly bi-Maxwellian, but instead simply
asymmetric to an extent that it is significantly better represented
by a bi-Maxwellian model than a symmetric Maxwellian
one.

     The $(\lambda_{axis},b_{axis})$ direction implied by {\em IBEX}
may not be very close to the $B_{ISM}$ direction, but it is only
$18.7^{\circ}$ from the He flow direction, $(\lambda_{flow},b_{flow})$,
as shown explicitly in Figures~3 and 11.  A direction picked at random
has only a 5.3\% chance of being this close to
$(\lambda_{flow},b_{flow})$, so the proximity of
$(\lambda_{axis},b_{axis})$ and $(\lambda_{flow},b_{flow})$ is
likely not accidental.  The heliocentric rest frame in which
$(\lambda_{flow},b_{flow})$ is computed has no relevance for the
VLISM in general, so the proximity between $(\lambda_{axis},b_{axis})$
and $(\lambda_{flow},b_{flow})$ implies that the VDF asymmetry
that we are sensing is not intrinsic to the VLISM, but is instead
induced in the VDF somewhere in the outer heliosphere.

     One possible cause of the VDF asymmetry is suggested by the
existence of the secondary He neutrals discovered by {\em IBEX},
which are likely produced by charge exchange in the outer heliosheath
\citep{mak14,mak16}.  Given that for He the primary
charge exchange reaction should be
${\rm He^{0}} + {\rm He}^{+}\rightarrow {\rm He}^{+} + {\rm He^{0}}$,
the primary He flow population that we are studying here represents
the neutral charge exchange partners for the ${\rm He}^{+}$ that
becomes the secondary neutral population after charge exchange.  This
outer heliosheath charge exchange therefore represents a loss process
for the primary neutral He flow.  If the charge exchange losses are
isotropic throughout the VDF, then there would be no effect on the VDF
shape.  But if the charge exchange losses are preferentially extracted
from certain parts of the VDF, then this would introduce asymmetries
into the VDF, which could be what we are detecting.

     Accounting for the secondary ISN He population itself remains a
concern as well.  Charge exchange adds a population to the overall
He VDF that is generally slower and hotter than the primary ISN He,
but subtracting the secondary He signal from the data is very
difficult.  Our subtraction is based on an analysis that treats the
secondaries as a homogeneous, laminar, Maxwellian flow from
infinity \citep{mak16}, much as the primary neutrals are
generally treated.  But this is a poor approximation for the secondary
population, which is created in a region of divergent flow in the
outer heliosheath, a region that is likely not very homogeneous.

     Additional theoretical work is needed to assess whether the outer
heliosheath charge exchange involving He should be expected to
introduce VDF asymmetries into the primary ISN He population consistent
with those we infer here.  The suggestion of non-Maxwellian
ISN VDFs also provides motivation for future investigations with
more advanced observing capabilities, such as those that will be
availabe on the proposed {\em Interstellar Mapping and Acceleration Probe}
({\em IMAP}) mission.

\section{Assuming a Kappa Distribution}

     In addition to the bi-Maxwellian VDF assumption discussed in
the previous section, we also experiment with a kappa distribution.
A kappa distribution has been a common way to describe space plasmas
that are not in equilibrium, such as the solar wind
\citep{vp10,gl12}.  Observational
evidence for turbulence exists in the ISM \citep{sr08},
and this turbulence could in principle lead to VDFs that are best
described with kappa distributions \citep{laf06}.

     A kappa distribution can be expressed as
\begin{equation}
F({\bf v})=\frac{n\Gamma(\kappa+1)}{w_o^3\pi^{3/2}
  \kappa^{3/2}\Gamma(\kappa-1/2)}
  \left(1+\frac{({\bf v}-{\bf U})^2}{\kappa w_o^2}\right)^{-\kappa-1},
\end{equation}
where
\begin{equation}
w_o=\sqrt{\frac{2kT}{m}\frac{\kappa-3/2}{\kappa}},
\end{equation}
with $T$ defined in the limit $\kappa\rightarrow\infty$.  This
distribution has a thermal core and a power law tail, reducing to a
Maxwellian as $\kappa\rightarrow\infty$.  The lower limit for
$\kappa$ is $\kappa=1.5$, where the kappa distribution
becomes ill-defined.

     Replacing a Maxwellian with a kappa distribution and reanalyzing
the {\em IBEX} and {\em Ulysses} data is significantly simpler than
the bi-Maxwellian analysis in Section~2, because the kappa
distribution only introduces one extra free parameter ($\kappa$),
compared to the three introduced by the bi-Maxwellian assumption.
The fitting process is therefore relatively straightforward.

     However, for {\em IBEX} we believe it is necessary to expand
the range of orbits and latitudes considered in order to properly
test the kappa distribution.  This is because the kappa distribution
is most distinct from a Maxwellian in its extended power law tail.
Thus, in addition to the orbit selection illustrated in Figure~5,
which was originally defined by \citet{mb15}, we add
up to two earlier orbits and up to two later orbits for each
considered year (2011--2014), with the precise additions each year
depending on the availability of good time intervals for each orbit.
We also increase the latitude range considered
from $\pm 25^{\circ}$ to $\pm 45^{\circ}$ from the ecliptic.

\begin{figure}[t]
\plotfiddle{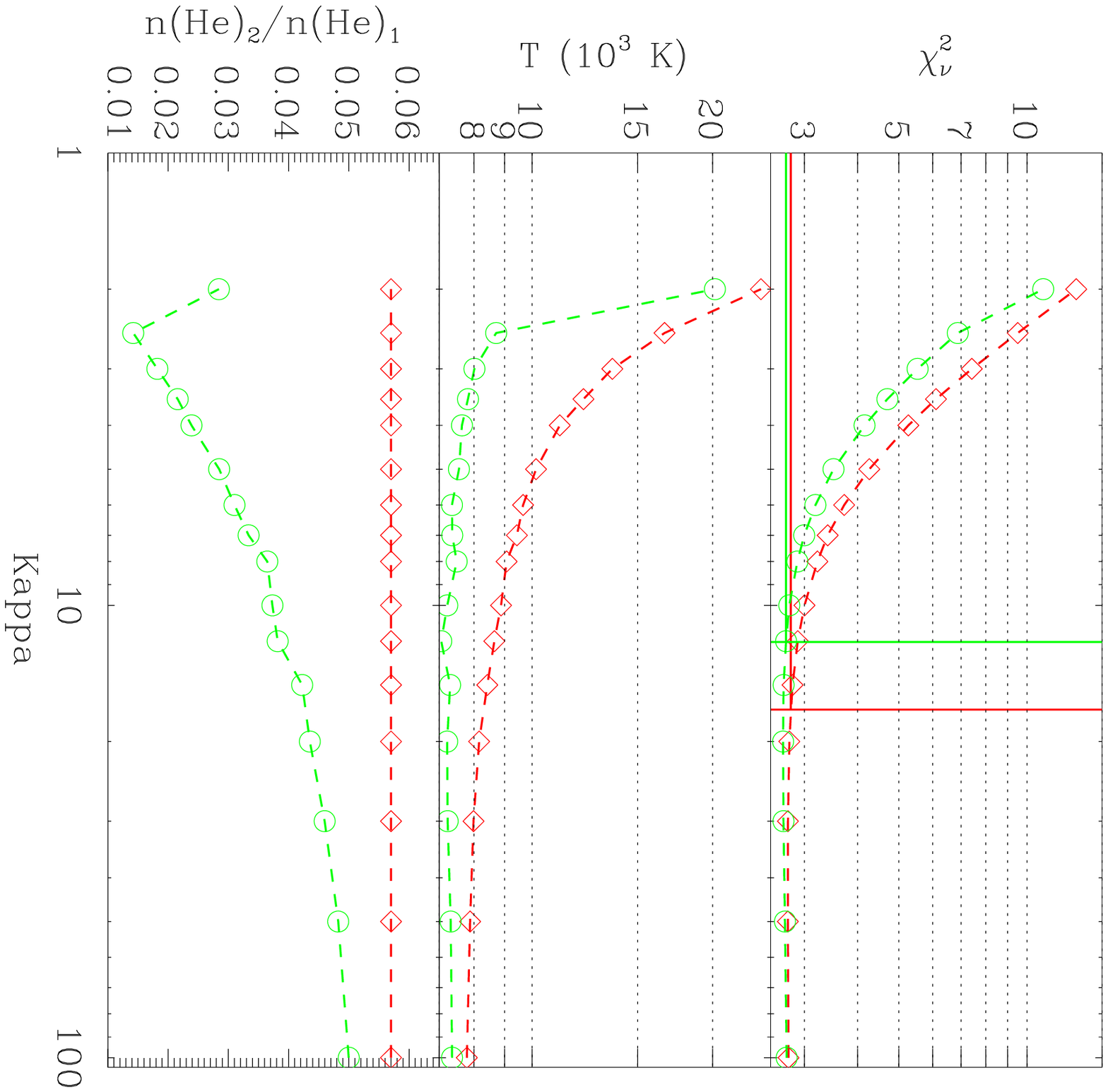}{3.7in}{90}{55}{55}{220}{-35}
\caption{Variations of $\chi^2_{\nu}$, best-fit temperature $T$,
  and secondary-to-primary neutral He ratio with assumed
  $\kappa$ in fits to {\em IBEX} data, assuming a
  kappa distribution for the He VDF.  Two sequences of fits
  are performed.  The sequence in red fixes the secondary
  contribution to be the  $n({\rm He})_2/n({\rm He})_1=0.057$
  value measured by Kubiak et al.\ (2016), while the sequence in
  green allows the $n({\rm He})_2/n({\rm He})_1$ ratio to vary in the
  fits.  In the $\chi^2_{\nu}$ panel, the horizontal solid
  lines are $3\sigma$ thresholds defining the boundary between
  acceptable and unacceptable fits.  The vertical solid lines are
  where these thresholds intercept the observed
  $\chi^2_{\nu}$-vs-$\kappa$ curves, defining lower limits for
  $\kappa$.}
\end{figure}
     Analogous to the bi-Maxwellian analysis in Figure~4, we
measure the $\kappa$ parameter and its uncertainty by conducting
a sequence of fits with different assumed $\kappa$ values held
constant, but all other parameters allowed to vary.  Results
are shown in red in the top panel of Figure~12.
We quickly conclude that the kappa
distribution does not provide an improvement in fits to the
data compared to a Maxwellian, as the best fits are for the
Maxwellian-like high-$\kappa$ models, with the $\chi^2_{\nu}$
value increasing greatly as $\kappa$ is decreased.  Using a
$3\sigma$ $\Delta\chi^2$ threshold, as in section~2, we find
an upper limit of $\kappa>17.0$ based on this initial analysis.
Figure~12 also illustrates the variation of best-fit temperature
as a function of $\kappa$, with $T$ decreasing monotonically
as $\kappa$ is increased.  This relation is very similar
to that found by \citet{ps19}, who have also
investigated the implications of a kappa distribution assumption
for analyses of {\em IBEX} data.

     In order to illustrate the problems with the low-$\kappa$
fits, Figure~5 shows the residuals of the best $\kappa=3$ fit.
There are clear, systematic discrepancies that are consistent
from year to year.  For example, for the last orbits of each
year shown in Figure~5 (116, 158a, 197b, and 238a), the
$\kappa=3$ fit systematically underpredicts fluxes near peak
center at latitudes in the range $[0^{\circ},10^{\circ}]$, and
overpredicts fluxes in the near wings, at $[-15^{\circ},-5^{\circ}]$
and $[15^{\circ},25^{\circ}]$.  This is clear evidence that the
extended power law nature of the wings of the low-$\kappa$ VDFs is not
consistent with the data.

     Perhaps the biggest systematic uncertainty in the
kappa distribution analysis concerns the effects of the secondary
He neutrals.  The hotter, slower secondary He population yields
a signal observed by {\em IBEX} over a broad distribution of
latitudes and orbits, stretching well beyond the orbits where
the primary He neutrals dominate.  The secondary component
therefore acts as a sort of broad background under the primary
component, the effects of which mimic the effects that should
be produced by a kappa distribution with broad power law wings.
We have to consider the possibility that when we correct for the
secondary neutrals (see Section~2.1), some of the counts that
we are assigning to the secondaries may in fact
be primary neutrals in the wings of a kappa distribution.

     For this reason, we perform a second {\em IBEX} analysis in
which we still assume the \citet{mak16} flow parameters
to correct for the secondaries, but we allow the contribution
fraction of the secondary population to vary from the
$n({\rm He})_2/n({\rm He})_1=0.057$ ratio measured by \citet{mak16}.
The results are shown in green in Figure~12.
Unlike for the original fit sequence, for this new sequence
$T$ does not initially increase much as $\kappa$ is decreased
from high values.  Instead, the now-variable
$n({\rm He})_2/n({\rm He})_1$ ratio decreases.  This essentially
frees up counts in the wings of the primary He component
for the wings of the low-$\kappa$ models to account for instead,
allowing those models to fit the data better than they did before.
Thus, the $\chi^2_{\nu}$ values decrease significantly.
Nevertheless, this experiment does not change the basic
conclusion that a kappa distribution does not provide a
statistically preferable fit to the data compared to a Maxwellian,
and the lower limit for $\kappa$ only decreases to $\kappa>12.1$.
This more conservative lower limit is what we choose
to quote as our final result.

     It should be noted that if we considered even earlier orbits
for each year, where the secondary neutrals completely dominate
the {\em IBEX} signal, the $\chi^2_{\nu}$ improvement for the
variable $n({\rm He})_2/n({\rm He})_1$ model in Figure~12 would
likely decrease, as the low $n({\rm He})_2/n({\rm He})_1$
values would probably lead to a failure to account for the
observed counts in those earlier orbits.  However, this only
emphasizes that our approach for assessing the systematic errors
introduced by uncertainties in the secondary He contribution
should be a fairly conservative one.

\begin{figure}[t]
\plotfiddle{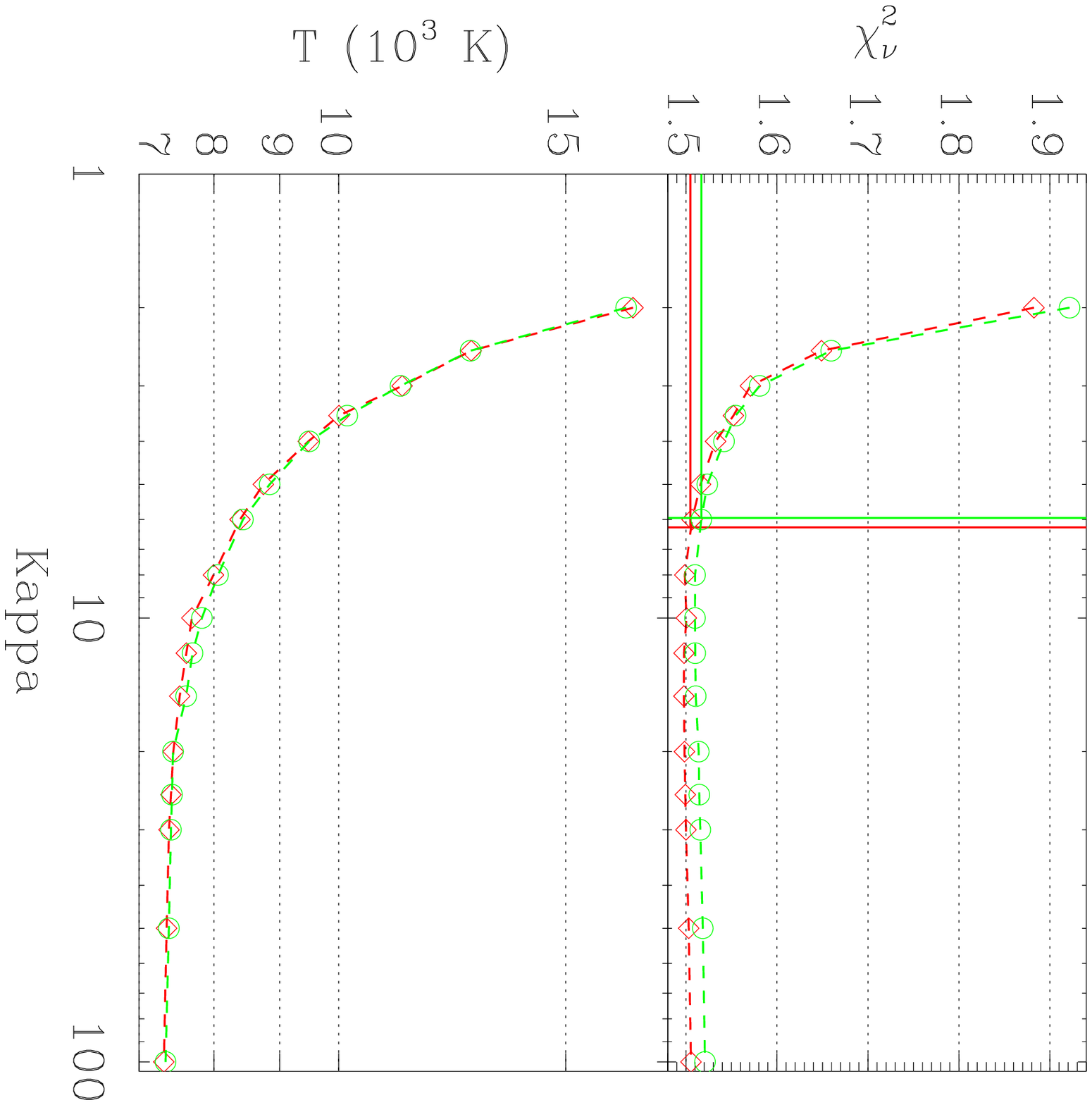}{3.7in}{90}{55}{55}{220}{-35}
\caption{Variations of $\chi^2_{\nu}$ and best-fit temperature $T$
  with assumed $\kappa$ in fits to {\em Ulysses} data assuming a
  kappa distribution for the He VDF.  Two sequences of fits
  are performed.  The sequence in red includes a correction for
  secondaries based on Wood et al.\ (2017), while the sequence in
  green does not.  In the $\chi^2_{\nu}$ panel, the horizontal solid
  lines are $3\sigma$ thresholds defining the boundary between
  acceptable and unacceptable fits.  The vertical solid lines are
  where these thresholds intercept the observed
  $\chi^2_{\nu}$-vs-$\kappa$ curves, defining lower limits for
  $\kappa$.}
\end{figure}
     Finally, we also repeat our analysis of the {\em Ulysses}
data, but assuming a kappa distribution instead of a Maxwellian
or bi-Maxwellian.  Initial results are shown in red in Figure~13.
We once again find no evidence that a kappa distribution
yields better fits than a simple Maxwellian.  We find a lower
limit of $\kappa>6.3$.

     Analogous to the {\em IBEX} analysis, we assess the
systematic errors introduced into the {\em Ulysses} analysis by
uncertainties in the He secondary subtraction.  For {\em Ulysses},
we address this by conducting a second sequence of fits in which
we simply do not correct for secondaries at all.
In principle, this should free up extra counts in the wings
of the observed He beam for the kappa distributions to account
for, allowing low-$\kappa$ models to fit the data better.
In practice, however, the change in $\chi^2_{\nu}$ is
minimal.  The results, shown in green in Figure~13, yield only
a tiny downward adjustment to the $\kappa$ lower limit,
$\kappa>6.0$, which is nevertheless what we choose to
report as our final measurement.  Given the lower S/N of the
{\em Ulysses} data relative to {\em IBEX}, it is not a surprise
that its limits on $\kappa$ are less restrictive.

     We find no evidence that a kappa distribution provides
better fits to either the {\em IBEX} or {\em Ulysses} data, compared
to a simple Maxwellian VDF.  This suggests that the VLISM
may be very close to thermal equilibrium, despite the apparent
presence of turbulence.  Although densities in the VLISM
are very low, and collisional timescales therefore very long, the
perturbation timescale in the VLISM may also be very long,
possibly providing sufficient time for collisional equilibrium
to be reached.  This would happen first for the plasma
component of the VLISM, which could then imprint its properties
onto the neutral component through charge exchange.

\section{Summary}

     We have reanalyzed {\em IBEX} and {\em Ulysses} observations
of interstellar neutral He atoms flowing through the inner solar
system.  Our primary goal has been to assess whether fits to the
data can be improved by assuming non-Maxwellian VDFs for the
flow, as opposed to the usual Maxwellian assumption.  Our findings
are summarized as follows:
\begin{description}
\item[1.] We first experiment with a bi-Maxwellian VDF.  In
  fits to {\em IBEX} data from 2011--2014 we find a very
  clear preferred fit with $T_{\perp}/T_{\parallel}=0.62\pm 0.11$
  oriented about an axis with ecliptic coordinates
  $(\lambda_{axis},b_{axis})=(57.2^{\circ}\pm 8.9^{\circ},
  -1.6^{\circ}\pm 5.9^{\circ})$.
  This result seems fairly robust to most assumptions made
  in the analysis.  However, we find some evidence that the
  $\lambda_{axis}$ parameter is curiously dependent on assumptions
  about the energy dependence of particle detection.  We also
  find some sensitivity of our results
  to the assumed photoionization loss rates, though it is questionable
  whether our assumed rates could be off by enough for this to be a
  major issue.  The
  $(\lambda_{axis},b_{axis})$ location right on the ecliptic is also
  cause for concern, as this is in {\em IBEX}'s orbital
  plane around the Sun.  This could in principle be
  indicative of some unknown systematic error affecting the
  {\em IBEX} data or the assumptions involved in our analysis of it.
\item[2.] A bi-Maxwellian analysis is also performed on
  {\em Ulysses} data.  The lower S/N of the data leads to
  the existence of two rather large, separate $(\lambda_{axis},b_{axis})$
  regions of acceptable fit, but one of them is consistent
  with the {\em IBEX} result, and it also has $T_{\perp}/T_{\parallel}<1$.
  Thus, the {\em Ulysses} analysis provides independent support
  for the best-fit {\em IBEX} bi-Maxwellian solution, and unlike
  {\em IBEX}, the {\em Ulysses} analysis will not be sensitive to
  photionization assumptions.
\item[3.] The axis of symmetry implied by the best bi-Maxwellian
  fit to the data, $(\lambda_{axis},b_{axis})$, is not very close to
  the $B_{ISM}$ direction, but it is only $18.7^{\circ}$ from the
  He flow direction in a heliocentric rest frame.  This implies that
  we are detecting VDF
  asymmetries that are introduced in the outer heliosphere rather
  than VDF asymmetries that are intrinsic to the VLISM.  Perhaps
  charge exchange processes in the outer heliosheath are
  responsible for these asymmetries.
\item[4.] We also experiment with a kappa distribution,
  but we find that a kappa distribution does not lead to better
  fits to the data for either {\em IBEX} or {\em Ulysses},
  compared to a Maxwellian.  Based on these analyses, we
  find lower limits of $\kappa>6.0$ and $\kappa>12.1$ based
  on {\em Ulysses} and {\em IBEX}, respectively.
\item[5.] The kappa distribution results emphasize that a
  simple Maxwellian VDF is actually quite successful at fitting
  the {\em IBEX} and {\em Ulysses} data, implying
  that the VLISM is close to thermal equilibrium.
\item[6.] Replacing the Maxwellian VDF with a bi-Maxwellian or
  kappa distribution does not lead to any significant change in the
  best-fit ISN He flow vector, for either {\em IBEX} or {\em Ulysses}.
\end{description}

\acknowledgments

Support for this project was provided by NASA award NNH16AC40I to
the Naval Research Laboratory.

\end{document}